\documentclass[a4paper,12pt,twoside]{article}
\usepackage{amsmath}
\usepackage{amssymb}
\usepackage{amsfonts}
\usepackage{amsthm}
\usepackage{mathrsfs}
\usepackage[pdfborder={0 0 0}]{hyperref}
\usepackage[utf8]{inputenc}
\usepackage{graphicx}

\newlength{\dinwidth}
\newlength{\dinmargin}
\newtheorem{theorem}{Theorem}[section]
\newtheorem{lemma}[theorem]{Lemma}
\newtheorem{proposition}[theorem]{Proposition}

\newtheorem{definition}[theorem]{Definition}

\numberwithin{equation}{section}

\usepackage{oldgerm}
\usepackage{color}

\newcommand{\Ibb}[1]{ {\rm I\ifmmode\mkern -3.6mu\else\kern -.2em\fi#1}}
\newcommand{\ibb}[1]{\leavevmode\hbox{\kern.3em\vrule
     height 1.2ex depth -.3ex width .2pt\kern-.3em\rm#1}}
\newcommand{\Cl}{{\ibb C}}
\newcommand{\Rl}{{\Ibb R}}
\newcommand{\Nl}{{\Ibb N}}
\newcommand{\Zl}{\mathbb{Z}}

\definecolor{lightgray}{rgb}{0.8,0.8,0.8}

\newcommand{\Om}{\Omega}
\newcommand{\om}{\omega}

\newcommand{\te}{\theta}

\newcommand{\la}{\lambda}
\newcommand{\La}{\Lambda}
\newcommand{\eps}{\varepsilon}

\newcommand{\Ba}{\mathcal{B}}

\newcommand{\Dc}{\mathcal{D}}
\newcommand{\A}{\mathcal{A}}
\newcommand{\R}{\mathcal{R}}
\newcommand{\B}{\mathcal{B}}

\newcommand{\M}{\mathcal{M}}

\newcommand{\Hil}{\mathcal{H}}

\newcommand{\DD}{\mathcal{D}}

\newcommand{\Pol}{\mathcal{P}}
\newcommand{\SF}{\mathcal{S}}
\newcommand{\SFreg}{\mathcal{S}_{\rm reg}}
\newcommand{\SFlim}{\mathcal{S}_{\rm lim}}

\newcommand{\Ss}{\mathscr{S}}   

\newcommand{\We}{\mathscr{W}}

\newcommand{\Mhat}{\widehat{\cal M}}

\newcommand{\fti}{\tilde{f}}

\newcommand{\ghat}{\hat{g}}
\newcommand{\fhat}{\hat{f}}

\newcommand{\fbar}{\overline{f}}

\newcommand{\PGpo}{\mathcal{P}_+^\uparrow}   

\newcommand{\PG}{\mathcal{P}}

\newcommand{\frS}{\textfrak{S}}

\def\bbeta{{\mbox{\boldmath{$\beta$}}}}

\newcommand{\Strip}{\mathrm{S}}

\newcommand{\OO}{O}

\newcommand{\supp}{\operatorname{\mathrm{supp}}}

\newcommand{\sh}{\mathrm{sh}}
\newcommand{\ch}{\mathrm{ch}}

\newcommand{\im}{\operatorname{\mathrm{Im}}}
\newcommand{\re}{\operatorname{\mathrm{Re}}}

\newcommand{\ot}{\otimes}

\newcommand{\tp}[1]{^{\otimes #1}}    

\newcommand{\zd}{z^{\dagger}}

\newcommand{\even}{\mathsf{e}}
\newcommand{\loc}{\mathsf{loc}}
\newcommand{\yd}{y^{\dagger}}
\newcommand{\hrskp}[2]{ \big\langle #1 ,\, #2 \big\rangle}
\newcommand{\email}[1]{\href{mailto:#1}{#1}}

\newcommand{\ri}{r}
\newcommand{\lef}{l}
\newcommand{\odd}{\mathsf{o}}
\definecolor{Red}{rgb}{1,0,0}
\definecolor{Green}{rgb}{0,0.65,0}
\title{Scaling limits of integrable quantum field theories}
\author{Henning Bostelmann\thanks{University of York, Department of Mathematics,
York YO10 5DD, United Kingdom.
E-mail: \email{henning.bostelmann@york.ac.uk}}
\and
Gandalf Lechner\thanks{Institute for Theoretical Physics, University of
Leipzig, Vor dem Hospitaltore 1, 04103 Leipzig, Germany.
E-mail:
\email{gandalf.lechner@uni-leipzig.de} --- supported by FWF project
P22929--N16 ``Deformations of Quantum Field Theories"}
\and
Gerardo Morsella\thanks{University of Roma Tor Vergata, Department of
Mathematics, viale della Ricerca
Scientifica 1, I-00133 Roma, Italy, E-mail: \email{morsella@mat.uniroma2.it} ---
supported in part by ERC Advanced Grant
227458 ``Operator Algebras and Conformal Field Theory''}
}
\date{May 13, 2011}

\begin{document}
\maketitle
\begin{center}
  \emph{Dedicated to the memory of Claudio D'Antoni}
\end{center}
\vspace*{3mm}

\begin{abstract}
	Short distance scaling limits of a class of integrable models on
two-dimen\-sional Minkowski space are
considered in the algebraic framework of quantum field theory. Making use of the
wedge-local quantum fields generating
these models, it is shown that massless scaling limit theories exist, and
decompose into (twisted) tensor products of
chiral, translation-dilation covariant field theories. On the subspace which is
generated from the vacuum by the
observables localized in finite light ray intervals, this symmetry can be
extended to the M\"obius group. The structure
of the interval-localized algebras in the chiral models is discussed in two
explicit examples.
\end{abstract}

\section{Introduction}

In the analysis of quantum field theories, the information gained by computing
the ultraviolet scaling limit and
determining its properties is most relevant. Probably the most important
application of this principle in physics is
perturbative asymptotic freedom of QCD -- the key feature which led
to its general acceptance as the quantum field theory of strong interactions.

In view of its importance, several approaches to the computation of the scaling
limit have been developed,
adapted to different descriptions of quantum field theory. In the Lagrangian
framework, the requirement
that the physical amplitudes are independent of the arbitrary choice of the
distance (or energy) scale
at which the theory is renormalized, provides an equation for the dependence of
renormalized correlation function on
coupling constants and the renormalization scale (the Callan-Symanzik equation).
Using information from
perturbation theory, the coefficients of the equation (the $\beta$-functions)
can usually be determined, and the
scaling limit of correlation functions computed by solving it.

However, these methods
are of little help in cases in which perturbation theory is not reliable, or
where the theory is not defined in terms of
a Lagrangian at all. In order to circumvent these problems, a different
approach, based on the algebraic setting
of quantum field theory~\cite{Haag:1996}, has been proposed by Buchholz and
Verch~\cite{Buchholz:1995a}
and extended in~\cite{BostelmannDAntoniMorsella:2007,Bostelmann:2008hr}.
In this approach, one considers the \emph{scaling algebra}, i.e., the algebra
generated by functions $\lambda \mapsto
A_\lambda$ of the scaling parameter with values in the algebra of observables of
the theory, satisfying certain
specific phase space properties. The scaling limit is then obtained as the GNS
representation of
the scaling algebra induced by the scaling limit of the vacuum state on the
original algebra at finite
scales. Relying only on the knowledge of the observables of the theory, this
method is completely
model-independent, and it proved to be very useful in analyzing the scaling
limit of charged sectors,
and in providing an intrinsic definition of confined charge~\cite{Buc:quarks,
DMV:scaling_charges}. A study
of the relations with the Lagrangian approach can be found
in~\cite{BostelmannDAntoniMorsella:2007}.

In the present article we study scaling limits of a certain class of integrable
quantum field theories on
two-dimensional Minkowski space. It is interesting to note that two-dimensional
sigma models, which are integrable field
theories -- although not directly covered by our results -- share with QCD the
property of asymptotic freedom, as well
as several others (see e.g.~\cite{ZinnJustin:1989} and references therein). We
will study a simplified version of these:
At finite scale, the models we are interested in describe a single type of
scalar neutral Bosons of mass $m>0$, whose
collision theory is governed by a factorizing S-matrix. This means that the
particle number is conserved in each
scattering process and the $n$-body S-matrix factorizes into a product of
two-body S-matrices, cf.~the textbook and
review \cite{AbdallaAbdallaRothe:1991,Dorey:1998} and the references cited
therein. A prominent example of a model in
the considered class is the Sinh-Gordon model.

We are particularly interested in the connection between the long and short
distance regimes of such quantum field
theories, represented by the S-matrix on the one hand and the scaling limit on
the other hand. For the simplified
particle spectrum that we consider here, a factorizing S-matrix can be fully
characterized by a single complex-valued
function $S$, the so-called scattering function. It is therefore possible to
formulate these models in the spirit of
inverse scattering theory, taking a scattering function $S$ and a mass value
$m>0$ as an input. Such a setup is directly
related to the long distance regime, and will be more convenient for our
analysis than Euclidean perturbation theory
(see, for example, \cite{Froehlich:1975}), where the relation to the real time
S-matrix is quite indirect.

There exist different, complementary, approaches to the inverse scattering
problem. One such approach, known as the form
factor program, aims at computing $n$-point functions of local fields in terms
of so-called form factors, i.e., matrix
elements of field operators in scattering states \cite{Smirnov:1992}. But
despite many partial results known in the
literature \cite{BabujianKarowski:2004}, in this approach one usually runs into
the problem that the convergence of the
appearing infinite series cannot be controlled because of the complicated form
of local field operators
\cite{BabujianFoersterKarowski:2006}. Another approach is based on the
operator-algebraic framework of quantum field
theory and Tomita-Takesaki modular theory, and constructs the models in question
by an indirect procedure involving
auxiliary field operators with weakened localization properties. Instead of
being sharply localized at points in
space-time, these fields (``polarization-free generators"
\cite{BorchersBuchholzSchroer:2001})
are localized only in infinite wedge-shaped regions ("wedges"). This last
approach will be most convenient for our
purposes as it is closely connected to the S-matrix and does not rely on series
expansions with unknown convergence
properties. Starting from a scattering function $S$ and a mass $m>0$, a solution
to the inverse scattering problem has
been rigorously constructed in this setting \cite{Schroer:1997-1,
SchroerWiesbrock:2000-1,
Lechner:2003,BuchholzLechner:2004,Lechner:2008}. The main results of this
analysis will be recalled in Section
\ref{section:2dModels} in a manner adapted to scaling transformations.

The resulting models meet all standard requirements of algebraic quantum field
theory, and hence on abstract grounds, a
well-defined scaling limit in the sense of Buchholz and Verch exists. In
particular, the short distance regime is in
principle completely described by the initially chosen S-matrix. We will not
fully analyze the Buchholz-Verch limit
here, but choose a simplified construction in the same spirit.

For the limit theory, one has natural candidates: massless models with
factorizing scattering. These have been described
before in a thermodynamical context; see
e.g.~\cite{ZamolodchikovZamolodchikov:1992,FendleySaleur:1994}. Here, however,
we treat them as rigorously constructed quantum field theories on
two-dimensional Minkowski space, given in terms of
local algebras of observables.  These limit theories are interesting in their
own right as they provide non-trivial
covariant deformations of free field theories (see also \cite{Lechner:2011} for
higher-dimensional generalizations), and
still depend on the scattering function $S$ one started with. Furthermore, as
expected for a scaling limit
\cite{Bostelmann:2008hr}, they are dilation covariant and, as it turns out,
(extensions of) chiral nets. This distinguishes them from other
massless deformations of quantum field theories that have recently been
constructed in the algebraic framework, on
two-dimensional Minkowski space \cite{DybalskiTanomoto:2010} and Minkowski
half-space \cite{LongoRehren:2004,LongoWitten:2011}.

In this paper, we start to explore the relation between the scattering function
defining a massive model of the class
mentioned above, and the properties of the corresponding scaling limit.

The first step consists of computing the behavior of the scattering function
under scaling transformations, and to
determine the short distance structure of the $n$-point functions of the
wedge-local generators. This is done in Section
\ref{section:2dModels} and Section \ref{section:MassiveLimit}, respectively. As
expected, the mass vanishes in the short
distance limit, and we obtain a class of massless (local extensions of) chiral
quantum field theories. They are presented in Section
\ref{section:ChiralModels}. As we shall explain, their dependence on $S$ is
twofold: On the one hand, $S$ determines the
decomposition of the  two-dimensional massless generators into twisted or
untwisted tensor products of chiral fields on
the left and right light ray. On the other hand, the chiral components can be
generated by massless chiral quantum
fields which are localized on half-lines, similar to the massive situation. The
commutation relations of these fields
directly involve the scattering function $S$ in a manner very similar to the
two-
dimensional models at finite scale, despite the difference in mass and
space-time dimension.

The chiral subtheories always transform covariantly under a representation of
the translation-dilation-reflection group
of the light ray. Making use of modular theory, we will show that on a subspace
of the chiral Hilbert space, one can
always extend this affine symmetry by a conformal rotation to the M\"obius group
(Section \ref{section:Conformal}). This
conformal subspace is directly related to observables localized in finite
intervals on the light ray. But because our
construction is based on half\/line-local generator fields, such strictly
localized observables are derived quantities
here, and it is a non-trivial task to characterize them.

We obtain two results in this direction: First, we show that for certain
scattering functions, the local chiral
observables are fixed points under an additional $\Zl_2$-symmetry, which
restricts the conformal subspace. Second, we
investigate the models given by two simple example scattering functions in full
detail in Section
\ref{section:IsingModel}. In these examples, we find conformal nets with central
charge $c=1$ respectively
$c=\frac{1}{2}$ in the limit. This analysis also exemplifies that the scaling
limit of a Bosonic theory can be generated
by the energy-momentum tensor of a Fermi field.

Section \ref{section:conclusion} contains our conclusions and an account of
further work in progress.

\section{Two-dimensional integrable models}\label{section:2dModels}

In this section, we recall the structure of the quantum field theories we are
interested in. At finite scale, these
models describe a single species of scalar Bosons of mass $m\geq0$ on
two-dimensional Minkowski space. Scattering
processes of these particles are governed by a {\em factorizing S-matrix}
\cite{AbdallaAbdallaRothe:1991,
BabujianFoersterKarowski:2006, Dorey:1998}, i.e., in each collision process the
particle number and the momenta are
conserved, and the $n$-particle S-matrix factorizes into a product of
two-particle S-matrices. In this situation, the
S-matrix is determined by a single function $S$, called the {\em scattering
function}. Such a restricted form of the
collision operator is typical for completely integrable models
\cite{Iagolnitzer:1978}, which provide a rich class of
examples for factorizing S-matrices.

The family of model theories we consider is thus parametrized by the two data
$(m,S)$, where $m$ is a mass parameter and
$S$ a function with a number of properties specified below. Before recalling the
construction of these quantum field
theories, we define the space of parameters $(m,S)$ and investigate its scaling
properties. We will first consider the
case $m>0$, and then obtain the massless case $m=0$ in a suitable limit.

\subsection{Scaling limits of scattering
functions}\label{subsection:ScalingOfScatteringFucntions}

The defining properties of a scattering function $S$ can most conveniently be
expressed when treating $S$ as a function
of the rapidity $\te$ as the momentum space variable, which parametrizes the
upper mass shell with mass $m>0$ according
to
\begin{align}\label{pte}
	p_m(\te)
	:=
	m
	\left(
	\begin{array}{c}
		\cosh\te\\
		\sinh\te
	\end{array}
	\right)
	\,,\qquad\te\in\Rl\,.
\end{align}
Since Lorentz boosts are translations in the rapidity, the Lorentz invariant
scattering function depends only on
differences of rapidities. Writing
\begin{align}
	\Strip(a,b) := \{\zeta\in\Cl\,:\,a<{\rm Im}\,\zeta<b\}
\end{align}
for two real numbers $a<b$, and $\overline{\Strip(a,b)}$ for the closed strip,
the family of all scattering functions
and two important subfamilies are defined as follows.

\begin{definition}{\bf (Scattering
functions)}\label{definition:ScatteringFunctions}\\
\vspace*{-7mm}
\begin{enumerate}
 \item A \emph{scattering function} is a bounded and continuous function
$S:\overline{\Strip(0,\pi)}\to\Cl$ which is
analytic in the interior of this strip and satisfies for $\te\in\Rl$,
\begin{equation}\label{eq:ScatteringFunctionRelations}
	\overline{S(\te)}		=	S(\te)^{-1}	=
S(\te+i\pi)		=	S(-\te)\,.
\end{equation}
The family of all scattering functions is denoted $\SF$.
\item A scattering function $S\in\SF$ is called \emph{regular} if there exists
$\kappa>0$ such that $S$ continues to a
bounded analytic function in the strip $\Strip(-\kappa,\pi+\kappa)$. The
subfamily of all regular scattering functions
is denoted $\SFreg\subset\SF$.
\item A regular scattering function $S\in\SFreg$ is called a \emph{scattering
function with limit} if the two limits
$\lim_{\te\to\infty}S(\te)$ and $\lim_{\te\to-\infty}S(\te)$ exist. The family
of all scattering functions with limit is
denoted $\SFlim\subset\SFreg\subset\SF$.
\end{enumerate}
\end{definition}

The equations \eqref{eq:ScatteringFunctionRelations} express the unitarity,
crossing symmetry, and hermitian analyticity
of the factorizing S-matrix corresponding to $S$. For a discussion of these
standard properties, we refer to the
textbooks and reviews \cite{AbdallaAbdallaRothe:1991, BabujianKarowski:2004,
Iagolnitzer:1993,Smirnov:1992, Dorey:1998}.
The regularity assumption in part b) of Definition
\ref{definition:ScatteringFunctions} comes from the fact that for
each regular $S\in\SFreg$, a corresponding quantum field theoretic model is
known to exist \cite{Lechner:2008}, whereas
for non-regular scattering functions, this is not known. Particular examples of
regular scattering functions (with
limit) are the constant functions $S_{\rm free}(\te)=1$ and $S_{\rm
Ising}(\te)=-1$, corresponding to the
interaction-free theory and the Ising model, respectively, and the scattering
function of the Sinh-Gordon model with
coupling constant $g\in\Rl$ \cite{ArinshteinFateevZamolodchikov:1979},
\begin{align}\label{eq:SinhGordonScatteringFunction}
	S_{\rm ShG}(\te)
	:=
	\frac{\sinh\te-i\sin\frac{\pi g^2}{4\pi+g^2}}{\sinh\te+i\sin\frac{\pi
g^2}{4\pi+g^2}}
	\,.
\end{align}
The additional assumption in part c) of the above definition, concerning the
existence of limits of scattering
functions, is relevant in the context of scaling limits: If distances in
Minkowski space are scaled according to
$x\to\la x$, and Planck's unit of action $\hbar$ is kept fixed, momenta have to
be rescaled according to
$p\to\la^{-1}p$. So rapidities scale like
$\te=\sinh^{-1}\frac{p}{m}\to\sinh^{-1}\frac{p}{\la m}$ and converge to
$\pm\infty$ for $\la\to0$. Looking at the example of the Sinh-Gordon scattering
function
\eqref{eq:SinhGordonScatteringFunction}, where the coupling constant is
dimensionless and therefore does not scale with
$\la$, we see that the only dependence of $S(\te_1-\te_2)$ on the scale $\la$ is
via the scale dependence of
$\te_1,\te_2$. Hence for $S(\te_1-\te_2)$ to have a scaling limit as $\la\to0$,
we need to require the existence of the
limits as in part c).

An explicit characterization of such functions is given in the following
proposition.

\begin{proposition}{\bf (Scattering functions with
limits)}\label{proposition:ScatteringFunctionsWithLimits}\\
\vspace*{-7mm}
\begin{enumerate}
	\item The set $\SFlim$ of scattering functions with limits consists
precisely of the functions
 \begin{align}\label{eq:GeneralFormOfScatteringFunctionWithLimit}
	S(\zeta)
  	&=
  	\eps\cdot \prod_{k=1}^N
  	\frac{\sinh\zeta-\sinh b_k}{\sinh\zeta+\sinh b_k}
  	\,,\qquad
  	\zeta\in\overline{\Strip(0,\pi)}\,,
 	\end{align}
where $\eps=\pm1$, $N\in\Nl_0$, and $\{b_1,...,b_N\}$ is a set of complex
numbers in the strip $0< \im b_1,...,\im
b_N\leq\frac{\pi}{2}$, such that with each $b_k$ (counted according to
multiplicity) also $-\overline{b_k}$ is contained
in $\{b_1,...,b_N\}$.
\item For each $S\in\SFlim$, the two limits
$S(\infty):=\lim_{\te\to\infty}S(\te)=\lim_{\te\to-\infty}S(\te)$ coincide
and are equal to $\pm1$, {\it i.e.}, $\SFlim$ is the disjoint union of the sets
\begin{align}\label{eq:ScatteringFunctionsWithLimit+-1}
	\SFlim^\pm
	:=
	\{S\in\SFlim:\lim\limits_{\te\to\infty}S(\te)=\lim\limits_{\te\to-\infty}
S(\te)=\pm1\}\,.
\end{align}
\end{enumerate}
\end{proposition}
\begin{proof}
a) Each factor $s_{b_k}:\zeta\mapsto\pm(\sinh\zeta-\sinh b_k)(\sinh\zeta+\sinh
b_k)^{-1}$ satisfies
$s_{b_k}(-\zeta)=s_{b_k}(\zeta+i\pi)=s_{b_k}(\zeta)^{-1}=\overline{s_{-\overline
{b_k}}(\zeta)}$ for $\zeta\in\Rl$. Given
any sufficiently small $\delta>0$, the function $s_{b_k}$ is analytic and
bounded in the strip
$\Strip(-\im{b_k}+\delta,\pi+\im{b_k}-\delta)\supset \Strip(0,\pi)$. Because the
product
\eqref{eq:GeneralFormOfScatteringFunctionWithLimit} is finite, it follows that
$S$ is analytic and bounded in the strip
$\Strip(-\kappa,\pi+\kappa)$ for some $\kappa>0$. Furthermore, the last two
equations in
\eqref{eq:ScatteringFunctionRelations} hold for $S$ because they hold for each
factor $s_{b_k}$. The first equation in
\eqref{eq:ScatteringFunctionRelations} holds because of
$s_{b_k}(\zeta)^{-1}=\overline{s_{-\overline{b_k}}(\zeta)}$ and
the assumed invariance of $\{b_1,...,b_N\}$ under $b_k\to-\overline{b_k}$. Hence
each $S$ of the form
\eqref{eq:GeneralFormOfScatteringFunctionWithLimit} is a regular
scattering function. As $\te\to\pm\infty$, we clearly have $S(\te)\to\eps$,
which shows $S\in\SFlim$.

Now we pick some arbitrary $S\in\SFlim$ and show that it is of the form
\eqref{eq:GeneralFormOfScatteringFunctionWithLimit}. As a regular scattering
function, $S$ is bounded and analytic in a
strip $\Strip(-\kappa,\pi+\kappa)$ for some $\kappa>0$, and since $S\in\SFlim$,
we have a limit value $\eps\in\Cl$ such
that $S(\te)\to\eps$ as $\te\to\infty$. These properties imply that
$S(\te+i\la)\to \eps$ as $\te\to\infty$, uniformly
in $\la\in[0,\pi]$ \cite[p.~170]{Titchmarsh:1939}. In view of
$S(\te+i\pi)=S(-\te)$, we also have $S(\te+i\la)\to\eps$
for $\te\to-\infty$. In particular, the two limits
$\lim_{\te\to\pm\infty}S(\te)$ along the real line coincide.

Since $S$ has unit modulus on the real line
\eqref{eq:ScatteringFunctionRelations}, we have $|\eps|=1$, and because of
the uniform limit $S(\zeta)\to\eps$ as $\re(\zeta)\to\pm\infty$, we find $c>0$
such that $|\re(\zeta_0)|\leq c$ for all
zeros $\zeta_0$ of $S$. Taking into account that $S$ is continuous on the closed
strip $\overline{\Strip(0,\pi)}$, and
of modulus 1 on its boundary, we conclude that it has only finitely many zeros
in $\Strip(0,\pi)$.

Let us denote by $b_1,...,b_N$ those zeros of $S$ whose imaginary parts $\la$
satisfy $0<\la\leq\frac{\pi}{2}$. These
zeros come in pairs $\{b_k,-\overline{b_k}\}$ because of
\eqref{eq:ScatteringFunctionRelations}, and there also exist
corresponding zeros $i\pi-b_k, i\pi+\overline{b_k}$ in the upper half of the
strip. Now consider the product
\begin{align*}
	B(\zeta):= \eps\cdot  \prod_{k=1}^N\frac{\sinh\zeta-\sinh
b_k}{\sinh\zeta+\sinh b_k}\,,
\end{align*}
which is a regular scattering function $B\in\SFreg$, of the form specified in
\eqref{eq:GeneralFormOfScatteringFunctionWithLimit}. Since $B$ has precisely the
same zeros as $S$ in $\Strip(0,\pi)$,
and $B(\te+i\la)\to \eps$ for $\te\to\pm\infty$, also $F:=S\cdot B^{-1}$ belongs
to $\SF$.

By construction, $F$ has no zeros in $\Strip(0,\pi)$, and $F(\te+i\la)$
converges to $1$ for $\te\to\pm\infty$,
uniformly in $0\leq\la\leq\pi$. As $F$ is continuous on
$\overline{\Strip(0,\pi)}$ and of modulus 1 on the boundary of
this strip, it is bounded from above and below, i.e., there exists $K>0$ such
that $K<|F(\zeta)|\leq 1$,
$\zeta\in\overline{\Strip(0,\pi)}$. But any scattering function, and in
particular $F$, can be meromorphically continued
to $\Strip(-\pi,\pi)$ by the equations
\eqref{eq:ScatteringFunctionRelations}. In fact, this continuation is given
by
\begin{align}\label{eq:ScatteringFunctionsRelationsContinued}
 	F(-\zeta)
 	=
 	F(\zeta)^{-1}
 	\,,\qquad
 	\zeta\in\overline{\Strip(0,\pi)}\,,
\end{align}
and as $F$ has no zeros in $\overline{\Strip(0,\pi)}$, it is actually an
analytic continuation for this special
scattering function. In view of the boundedness of $F$ on
$\overline{\Strip(0,\pi)}$, there also holds
$|F(\zeta)|<K^{-1}<\infty$ for all $\zeta\in\overline{\Strip(-\pi,\pi)}$. Taking
$\zeta=-\te+i\pi$, $\te\in\Rl$, Eqs.
\eqref{eq:ScatteringFunctionsRelationsContinued} and
\eqref{eq:ScatteringFunctionRelations} give
\begin{align*}
 	F(\te-i\pi)
 	=
 	F(i\pi-\te)^{-1}
 	=
 	F(\te)^{-1}
 	=
 	F(\te+i\pi)
 	\,,\qquad
 	\te\in\Rl\,,
\end{align*}
i.e., $F$ continues to a $(2\pi i)$-periodic, entire function which in view of
the above argument is bounded and hence
constant. Thus $F(\te)=\lim_{\te\to\infty}F(\te)=1$, and we arrive at the
claimed representation
\eqref{eq:GeneralFormOfScatteringFunctionWithLimit} for $S$, namely $S=F\cdot
B=B$.

b) The identity of the limits $\lim_{\te\to\pm\infty}S(\te)$ has been shown
above, and can also be seen directly from
\eqref{eq:GeneralFormOfScatteringFunctionWithLimit}. Also the fact that these
limits can take only the values $\pm1$ is
clear from \eqref{eq:GeneralFormOfScatteringFunctionWithLimit}.
\end{proof}

As a preparation for the scaling limit of quantum fields, we now compute which
effect a space-time scaling $x\to\la x$
has on a scattering function with limit. As usual, such a limit involves taking
the mass to zero. To keep track of the
mass scale, we will use momentum variables with explicit mass dependence instead
of the rapidity. For spatial momenta
$p=m\sinh\te$, $q=m\sinh\te'$, we have
\begin{align*}
	\te-\te'
	&=
	\sinh^{-1}\frac{p}{m}-\sinh^{-1}\frac{q}{m}
	=
	\sinh^{-1}\left(\frac{p\,\om_q^m-q\om_p^m}{m^2}\right)
	\,,
\end{align*}
with the energies $\om_p^m:=(p^2+m^2)^{1/2}$, $\om_q^m:=(q^2+m^2)^{1/2}$.
Corresponding to any $m>0$, $S\in\SFlim$, we
therefore introduce the function $S_m:\Rl^2\to\Cl$,
\begin{align}\label{eq:ScatteringFunctionWithMass}
	S_m(p,q)
	:=
	S\left(\sinh^{-1}\left(\frac{p\,\om_q^m-q\om_p^m}{m^2}\right)\right)
 	\,,
\end{align}
which shows the mass dependence explicitly. Clearly, $S_m$ inherits many
properties from $S$, see
Eq.~\eqref{eq:ScatteringFunctionRelations}. For example, one has the symmetry
and scaling relations, for $p,q\in\Rl$,
\begin{align}
	S_m(q,p) \label{eq:SM-flip}
	&=
	S_m(p,q)^{-1}
	=
	\overline{S_m(p,q)}
	\,,\\
 	S_m(\la^{-1}p,\la^{-1}q)
 	&=
 	S_{\la m}(p,q)
 	\,,\qquad
 	\la>0\,.\label{eq:SM-scale}
\end{align}
The mass zero limit $S_0$ of $S_m$ can be computed in a straightforward manner.

\begin{lemma}\label{lemma:ScalingLimitOfScatteringFunction}
Let $S\in\SFlim^\pm$, and $m>0$. Then,  for $p,q\in\Rl$,
\begin{align}\label{eq:S0}
	S_0(p,q)
	:=
	\lim_{\la\to0}
	S_{\la m}(p,q)
	=
	\left\{
		\begin{array}{rcl}
		S(\log p-\log q)		&;& p>0,\,q>0\\
 		S(\log (-q)-\log (-p))		&;& p<0,\,q<0\\
	 	S(0)				&;&	p=q=0\\
		S(\infty)			&;&\text{otherwise}
	\end{array}
\right.
\quad .
\end{align}
\end{lemma}
\begin{proof}
For any $p,q\in\Rl$, we have
\begin{align}
	\lim_{\la\to0}(p\,\om^{\la m}_q-q\om^{\la m}_p)
	=
	p|q|-q|p|
	=
	\left\{
	\begin{array}{rcl}
		\pm 2pq&;&p\cdot q<0\\
 		0&;&p\cdot q\geq0
	\end{array}
	\right.
.
 \end{align}
 This implies $(\la m)^{-2}(p\om^{\la m}_q-q\om^{\la m}_p)\to\pm\infty$ for
$\la\to0$ if $p\cdot q<0$, and since
$S\in\SFlim^\pm$, we get $S_{\la m}(p,q)\to S(\infty)$ for this configuration of
momenta. In the case $p\cdot q\geq0$,
we use l'Hospital's rule to compute the limit,
\begin{align*}
	\lim_{\la\to0}\frac{p\,\om^{\la m}_q-q\om_p^{\la m}}{\la^2m^2}
	&=
	\lim_{\la\to0}
	\frac{\frac{p\la m^2}{\om^{\la m}_q}-\frac{q\la m^2}{\om_p^{\la
m}}}{2\la m^2}
	=
	\frac{1}{2}\lim_{\la\to0}
	\left(\frac{p}{\om^{\la m}_q}-\frac{q}{\om_p^{\la m}}\right)
	\\
	&=
	\left\{
	\begin{array}{rcl}
		0	&;& p=q=0\\
		\eps(p)\cdot\infty	&;& p\neq0,\;q= 0\\
   		-\eps(q)\cdot\infty	&;& p=0,\;q\neq 0\\
   		\frac{1}{2} \left(\frac{p}{|q|}-\frac{q}{|p|}\right) &;& p\cdot
q>0
 	\end{array}
	\right.
  	.
\end{align*}
Here $\eps(p)$, $\eps(q)$ denotes the sign of $p,q$, respectively. Evaluating
these expressions in $S\circ\sinh^{-1}$
\eqref{eq:ScatteringFunctionWithMass} gives the claimed result.
\end{proof}

Note that the limit $S_0$ is not independent of the scattering function $S$; in
fact, $S$ can be completely recovered
from $S_0$ \eqref{eq:S0}. This can be seen as an indication that the short
distance behavior of the $(m,S)$-model will
depend on $S$ (but not on $m$). The limit behavior of the scattering functions
will be used in the calculation of the
scaling limit of the field theory models discussed in the next section.

\subsection{Massive and massless models with factorizing
S-matrices}\label{subsection:2dmodels}

We now turn to the description of the family of quantum field theoretic models
we are interested in. Each model in this family is specified by two parameters,
a mass value $m\geq0$ and a scattering function $S\in\SFlim$ with limit.

Whereas the most frequently used setting for the discussion of such models is
the form factor program \cite{BabujianKarowski:2004}, their rigorous
construction was accomplished only recently with the help of operator-algebraic
techniques. The initial idea of this program is due to Schroer
\cite{Schroer:1997-1, Schroer:1999} and consists in constructing certain
auxiliary field operators depending on $(m,S)$. Despite their weaker than usual
localization, these fields can be used to define a strictly local, covariant
quantum field theory in an indirect manner. The details of this construction,
and the passage to algebras of strictly localized observables, was carried out
in \cite{Lechner:2003,BuchholzLechner:2004,Lechner:2007,Lechner:2008}. In
particular, it has been shown that for any choice of $(m,S)$, $m>0$,
$S\in\SFreg$, there exists a corresponding quantum field theory with the
factorizing S-matrix given by $S$ as its collision operator. In the following,
we will outline the structure of these models using a momentum space
formulation. For details and proofs, we refer to the articles cited above.

Fixing arbitrary $S\in\SFlim$ and $m\geq0$, the function $S_m$ is defined via
\eqref{eq:ScatteringFunctionWithMass} for
$m>0$ and via the limit \eqref{eq:S0} for $m=0$. Note that the zero mass
function $S_0$ \eqref{eq:S0} can be
discontinuous at $(0,0)$ if the signs of $S(0)$ and $S(\infty)$ are different,
but still satisfies the symmetry
relations \eqref{eq:SM-flip}.

Most of the objects introduced below depend on the choice of $S$, but since we
will work with a fixed scattering
function in the following, we do not reflect this dependence in our notation.
The mass dependence, on the other hand,
will always be written down explicitly.
\\
\\
Having fixed $(m,S)$, we first describe the Hilbert space on which the
$(m,S)$-model is constructed. Starting from the
single particle space $\Hil_{m,1}:=L^2(\Rl,dp/\om^m_p)$, the $n$-particle spaces
$\Hil_{m,n}$, $n>1$, are defined as
certain $S_m$-symmetrized subspaces of the $n$-fold tensor product
$\Hil_{m,1}^{\ot n}$. To this end, one introduces
unitaries $D_n(\tau_j)$, $j=1,...,n-1$, on $\Hil_{m,1}^{\ot n}$,
\begin{align}\label{def:Dn}
	(D_n(\tau_j)\Psi_n)(p_1,...,p_n)
 	:=
 	S_m(p_{j+1}, p_j)
 	\cdot
 	\Psi_n(p_1,...,p_{j+1},p_j,...,p_n)
 	\,.
\end{align}
Using \eqref{eq:SM-flip}, one checks that these operators generate a unitary
representation $D_n$ of the group $\frS_n$
of permutations of $n$ letters which represents the transposition exchanging $j$
and $j+1$ by $D_n(\tau_j)$. The
$n$-particle space $\Hil_{m,n}$ of the $(m,S)$-model is defined as the subspace
of $\Hil_{m,1}^{\ot n}$ of vectors
invariant under this representation. Explicitly, the orthogonal projection
$P_n:\Hil_{m,1}^{\ot n}\to\Hil_{m,n}$ has the
form
\begin{align}\label{def:Pn}
	(P_n\Psi_n)(p_1,...,p_n)
	&:=
	\frac{1}{n!}\sum_{\pi\in\frS_n}
	S_m^\pi(p_1,...,p_n)
	\cdot\Psi_n(p_{\pi(1)},...,p_{\pi(n)})
	\,,
	\\
	S_m^\pi(p_1,...,p_n)
	&:=
	\prod_{\substack{ 1 \leq l < r \leq n \\ \pi(l) > \pi(r)}}
	S_m(p_{\pi(l)}, p_{\pi(r)})
	\,.
\end{align}
Setting $\Hil_{m,0}:=\Cl$, the {\em $S_m$-symmetric Fock space} over
$\Hil_{m,1}$ is
\begin{align}
 	\Hil_{m}
 	:=
 	\bigoplus_{n=0}^\infty
 	\Hil_{m,n}
 	\,,
\end{align}
i.e., its vectors are sequences $\Psi=(\Psi_0,\Psi_1,\Psi_2,...\,)$, with
$\Psi_0\in\Cl$, $\Psi_n\in\Hil_{m,n}$,
$n\geq1$, such that $\|\Psi\|^2:=|\Psi_0|^2+\sum_{n=1}^\infty \int
\frac{dp_1}{\om^m_1}\cdots
\frac{dp_n}{\om^m_n}\,|\Psi_n(p_1,...,p_n)|^2<\infty$. Here and in the following
we use the shorthand notation
$\om_k^m=\om_{p_k}^m=(p_k^2+m^2)^{1/2}$.

On $\Hil_{m}$, there exists a strongly continuous (anti-)unitary positive
energy representation $U_m$ of the full
Poincar\'e group $\PG$. Denoting by $(x,\te)\in\PGpo$ proper orthochronous
transformations consisting of a boost with
rapidity $\te$ and a subsequent spacetime translation along
$x=(x_0,x_1)\in\Rl^2$, we set
\begin{align} \label{eq:2dTranslationDilation}
 	(U_m(x,\te)\Psi)_n(p_1,...,p_n)
 	:=
 	e^{i\sum_{j=1}^n (\om_j^m\,x_0 - p_jx_1)}
 	\cdot
 	\Psi_n(\te p_1,...,\te p_n)
 	\,,
\end{align}
where $\te p_j:=\cosh\te\cdot p_j-\sinh\te\cdot \om^m_j$, $j=1,...,n$. The
space-, time-, and spacetime-reflections
$j_1(x_0,x_1):=(x_0,-x_1)$, $j_0(x_0,x_1):=(-x_0,x_1)$ and $j:=j_0j_1$ are
represented as
\begin{align}
	(U_m(j_1)\Psi)_n(p_1,...,p_n)
 	&:=
 	\Psi_n(-p_n,...,-p_1)\,,
 	\\
 	(U_m(j_0)\Psi)_n(p_1,...,p_n)
 	&:=
	 \overline{\Psi_n(-p_1,...,-p_n)}
 	\,,
 	\\
 	(U_m(j)\Psi)_n(p_1,...,p_n)
 	&:=
 	\overline{\Psi_n(p_n,...,p_1)}
 	\,.
	\label{eq:Uj}
\end{align}
Clearly, all vectors in $\Hil_{m,1}$ are eigenvectors of the mass operator with
eigenvalue $m$, and the vector
$\Om_m:=1\oplus0\oplus0\oplus...\in\Hil_m$, invariant under $U_m$, represents
the vacuum state. The finite particle
number subspace of $\Hil_m$ is denoted $\DD_m$.
\\
\\
On $\DD_m$, there act creation and annihilation operators $z_{m}^\#(\varphi)$,
$\varphi\in\Hil_{m,1}$, defined as
\begin{align}
	\label{eq:ZM}
 	(z_{m}(\varphi)\Psi)_n(p_1,...,p_n)
 	&:=
 	\sqrt{n+1}\int\frac{dq}{\om^m_q}\,
 	\varphi(q)\,\Psi_{n+1}(q,p_1,...,p_n)
 	\,,
 	\\
 	\zd_{m}(\varphi)
 	&:=
 	z_{m}(\overline{\varphi})^*
	\;\Longleftrightarrow\;
 	(\zd_{m}(\varphi)\Psi)_n
 	=
 	\sqrt{n}P_n(\varphi\ot\Psi_{n-1})
 	\,.
	\label{eq:ZdM}
\end{align}
Because of the $S_m$-symmetrization properties of the vectors in $\Hil_m$, the
distributional kernels $z_{m}^\#(p)$,
$p\in\Rl$, related to the above operators by the formal integrals
$z_{m}^\#(\varphi)=\int\frac{dp}{\om^m_p}\,\varphi(p)z_{m}^\#(p)$, satisfy the
exchange relations of the
Zamolodchikov--Faddeev algebra \cite{ZamolodchikovZamolodchikov:1979,
Faddeev:1984},
\begin{align}
 	z_{m}(p)z_{m}(q)
 	&=
 	S_m(p, q)\,z_{m}(q) z_{m}(p)
 	\,,\label{eq:ZFzz}
 	\\
 	\zd_{m}(p)\zd_{m}(q)
 	&=
 	S_m(p, q)\,\zd_{m}(q) \zd_{m}(p)
 	\,,\label{eq:ZFzdzd}
 	\\
 	z_{m}(p)\zd_{m}(q)
 	&=
 	S_m(q, p)\,\zd_{m}(q) z_{m}(p)
 	+
 	\om_p^m\,\delta(p-q)\cdot \boldsymbol{1}_{\Hil_{m}}
 	\,.\label{eq:ZFzdz}
\end{align}

Having described the Hilbert space of the $(m,S)$-model, we now construct field
operators on it, and first introduce the
necessary test functions\footnote{We will use the symbol $\Ss(\Rl^n)$ for the
Schwartz space on $\Rl^n$. Given some set
$\OO\subset\Rl^n$, we also write $\Ss(\OO):=\{f\in\Ss(\Rl^n)\,:\,\supp
f\subset\overline{\OO}\}$ for its subspace
supported in $\OO$.}. For $f\in\Ss(\Rl^2)$ we write
\begin{align}\label{eq:fm}
 	f^{m\pm}(p)
 	:=
 	\frac{1}{2\pi}
 	\int d^2x\,f(x)\,e^{\pm i(\om^m_p,\,p)\cdot x}
\end{align}
for the restrictions of the Fourier transform of $f$ to the upper and lower mass
shell of mass $m\geq0$. For $m>0$, we
have $f^{m\pm}\in L^2(\Rl,dp/\om_p^m)$, and can therefore consider
$f^{m\pm}\in\Hil_{m,1}$ as a single particle vector.
For $m=0$, however, the measure $dp/\om_p^0=dp/|p|$ is divergent at $p=0$, and
therefore we can claim
$f^{0\pm}\in\Hil_{0,1}$ only if $f^{0,\pm}(0)=0$, i.e., if $f$ is the derivative
(w.r.t.\ $x_0$ or $x_1$) of another
test function. Bearing this remark in mind, we define a field operator
$\phi_{m}$ as
\begin{align}\label{eq:phi}
 	\phi_{m}(f)
 	:=
 	\zd_{m}(f^{m+}) + z_{m}(f^{m-})
 	\,.
\end{align}
For general $S$, this operator is unbounded, but always contains $\DD_m$ in its
domain and leaves this subspace
invariant. Furthermore, one can show that $\phi_m(f)$ is essentially
self-adjoint for real-valued $f$. Regarding its
field-theoretical properties, the field $\phi_m$ is a solution of the
Klein-Gordon equation with mass $m$, has the
Reeh-Schlieder property, and transforms covariantly under proper orthochronous
Poincar\'e transformations,
\begin{align}\label{eq:PhiMCovariance}
	U_m(x,\te)\phi_{m}(f)U_m(x,\te)^{-1}
	=
	\phi_{m}(f_{\te,x})
	\,,\qquad
	f_{\te,x}(y)
	=
	f(\La_\te^{-1}(y-x))
	\,.
\end{align}
Here $\La_\te=\left( \begin{smallmatrix} \ch\te\;\sh\te \\ \sh\te\;\ch\te
\end{smallmatrix} \right)$ denotes the Lorentz
boost with rapidity $\te$.

For positive mass, these properties have been established in
\cite{Lechner:2003}. For $m=0$, the proof carries over
without changes if restricting to derivative test functions, i.e., if
$f\in\Ss(\Rl^2)$ is assumed to be of the form
$f(x)=\partial g(x)/\partial x_k$, $g\in\Ss(\Rl^2)$, $k=0,1$.

Regarding locality, we first note that in the trivial case $S=1$, the field
$\phi_m$ coincides with the free scalar field of mass $m$, which is of
course point-local. For $S\neq1$, however, $\phi_{m}(x)$ is {\em not} localized
at the spacetime point $x\in\Rl^2$, i.e., in general
$[\phi_{m}(x),\phi_{m}(x')]\neq0$ for spacelike separated $x,x'\in\Rl^2$.
Moreover, the covariance property \eqref{eq:PhiMCovariance} does {\em not} hold
for the spacetime reflection $U_m(j)$ \eqref{eq:Uj} if $S\neq1$, i.e., the field
\begin{align}
	\phi_m'(f)
	:=
 	U_m(j)\phi_m(f^j)U_m(j)^{-1}
 	\,,\qquad
 	f^j(x):=\overline{f(-x)},
\end{align}
is different from $\phi_m$ in this case. Nonetheless, $\phi_m'$ shares many
properties with $\phi_m$, such as the domain
and essential self-adjointness, the covariant transformation behavior w.r.t.
proper orthochronous Poincar\'e
transformations \eqref{eq:PhiMCovariance}, and $\phi_m'$ is also a solution of
the Klein-Gordon equation with the
Reeh-Schlieder property. For the construction of a local quantum field theory
with scattering function $S\neq1$, one has
to make use of both fields, $\phi_m$ and $\phi_m'$, and exploit their relative
localization properties.

For the formulation of this relative localization, we first recall that the {\em
right wedge} is the causally complete
region
\begin{align}
	W_R
 	:=
 	\{x\in\Rl^2\,:\,x_1>|x_0|\}
 	\,,
\end{align}
and its causal complement is $W_R'=-W_R=:W_L$, the {\em left wedge}.

Given $m>0$ and $S\in\SFreg$, it has been shown in \cite{Lechner:2003} that the
two fields $\phi_{m}$, $\phi_m'$ are
{\em relatively wedge-local} to each other in the sense that
\begin{align}\label{eq:PhimPhimCommutator}
	[\phi_m(f),\,\phi_m'(g)]\Psi = 0
	\,,\quad
	\supp f\subset W_L,\,\supp g\subset W_R,
	\quad
	f,g\in\Ss(\Rl^2),\Psi\in\DD_m\,.
\end{align}
The proof of this fact relies on the analytic properties of $S_m$. As we saw in
Lemma
\ref{lemma:ScalingLimitOfScatteringFunction}, $S_0$ can even be  discontinuous,
and therefore one cannot directly employ
the analyticity arguments in the case $m=0$. However, using a splitting in
chiral components, we will see in Section
\ref{subsection:Splitting} that \eqref{eq:PhimPhimCommutator} is nonetheless
still valid in the massless situation.

Having collected sufficient information about the auxiliary fields $\phi_m$,
$\phi_m'$, one can pass to an operator-algebraic
formulation and consider the von Neumann algebras generated by them,
\begin{align}
	\M_{m}
 	&:=
 	\{e^{i\phi'_{m}(f)}\,:\,f\in\Ss(W_R)\,\text{ real}\}''
 	\,,\\
	\Mhat_{m}
	&:=
 	\{e^{i\phi_{m}(f)}\,:\,f\in\Ss(W_L) \,\text{ real}\}''
 	\,.
\end{align}
Using the relative localization and Reeh-Schlieder property of the fields
$\phi_m$, $\phi_m'$, one can show that $\M_m$ and $\Mhat_m$ commute,
and that the vacuum vector $\Om_m$ is cyclic and separating for both of them.
The
modular data of these algebras act geometrically as expected from the
Bisognano-Wichmann theorem \cite{BisognanoWichmann:1976}. In particular, the
modular conjugation $J$ of $(\M_m,\Om_m)$ coincides with the spacetime
reflection
$U_m(j)$ \eqref{eq:Uj}, and with this
information, it is easy to see that $\M_m$ and $\Mhat_m$ are actually
commutants of each other, $\Mhat_m=\M_m'$ \cite{BuchholzLechner:2004}.
Taking into account the transformation properties of the field $\phi_m'$, it
also
follows that
\begin{align*}
	U_m(x,\te)\M_m U_m(x,\te)^{-1} \subset \M_m
	\,,\qquad
	x\in W_R\,,\te\in\Rl\,.
\end{align*}
In view of these properties, one can consistently define von Neumann algebras of
observables localized in double cones (intersections of two opposite wedges).
For $y-x\in W_R$, one defines $\OO_{xy}:=(W_R+x)\cap (W_L+y)$ and
\begin{align}\label{def:AO}
	 \A_m(\OO_{xy})
 	:=
 	U_m(x,0)\M_m U_m(x,0)^{-1} \cap U_m(y,0)\M_m'U_m(y,0)^{-1}
	\,.
\end{align}
Algebras associated to arbitrary regions can then be defined by additivity. The
assignment $\OO\mapsto\A_m(\OO)$ of spacetime regions in $\Rl^2$ to observable
algebras in $\B(\Hil_m)$ is the definition of the $(m,S)$-model in the framework
of algebraic quantum field theory \cite{Haag:1996}. Its main properties are
summarized in the following theorem.

\begin{theorem}\label{theorem:MassiveNet}
 Let $m>0$ and $S\in\SFreg$. Then the map $\OO\mapsto\A_m(\OO)$ of double cones
in $\Rl^2$ to von Neumann algebras in
$\B(\Hil_m)$ has the following properties:
 \begin{enumerate}
	 \item Isotony: $\A_m(\OO_1)\subset\A_m(\OO_2)$ for double cones
$\OO_1\subset\OO_2$.
 	\item Locality:  $\A_m(\OO_1)\subset\A_m(\OO_2)'$ for double cones
$\OO_1\subset\OO_2'$.
 	\item Covariance: $U_m(g)\A_m(\OO)U_m(g)^{-1}=\A_m(g\OO)$ for
each Poincar\'e transformation $g\in\PG$ and each double cone $\OO$.
 	\item Reeh-Schlieder property: If $S(0)=1$, there exists $r_0>0$ such
that for all double cones $\OO$ which are
Poincar\'e equivalent to $W_R\cap(W_L+(0,r))$ with some $r>r_0$, there holds
$\overline{\A_m(\OO)\Om_m}=\Hil_m$. If
$S(0)=-1$, this cyclicity holds without restriction on the size of $\OO$.
 	\item Additivity: $\M_m$ coincides with the smallest von Neumann algebra
containing $\A_m(\OO)$ for all double
cones $\OO\subset W_R$.
 	\item Interaction: The collision operator of the quantum field theory
defined by the algebras \eqref{def:AO} is
the factorizing S-matrix with scattering function $S$.
 \end{enumerate}
Statements a)--c) also hold if $m=0$ and $S\in\SFlim$.
\end{theorem}
\pagebreak
The above list shows that the elements
of the algebra $\A_m(\OO)$ \eqref{def:AO} can consistently be interpreted as the
observables localized in $\OO$ of a local, covariant quantum field theory
complying with all standard assumptions. Furthermore, the last item shows that
the net $\A_m$ so constructed provides a solution to the inverse scattering
problem for the factorizing S-matrix given by $S$.

Statements d)--f) of Theorem \ref{theorem:MassiveNet} are only known to hold in
the massive case since an important tool
for their proof, the split property for wedges \cite{Muger:1998}, and the
closely related modular nuclearity condition
\cite{BuchholzDAntoniLongo:1990-1}, is not satisfied in the massless case. Thus
these properties might or might not be
valid in the mass zero limit. In Section \ref{section:IsingModel}, we will see
examples of scattering functions
$S\in\SFlim$ for both possibilities.

\section{Scaling limit of massive models}\label{section:MassiveLimit}

As a quantum field theory in the sense of Haag-Kastler \cite{Haag:1996}, the
models given by the nets $\A_m$ have a
well-defined scaling limit theory \cite{Buchholz:1995a}. However, for generic
scattering function, the local observables
of these models are given in a quite indirect manner as elements of an
intersection of two wedge algebras. On the other
hand, field operators localized in wedges are explicitly known, so that it is
not difficult to calculate their behavior
under scaling transformations.

To get an idea about the algebraic short distance limit of the $(m,S)$-models,
$m>0$, we can proceed in the following
way, inspired by the results in~\cite{BostelmannDAntoniMorsella:2007} about the
behavior of quantum fields under
scaling. We consider rescaled wedge-local field operators of the form
\begin{align}
 N_\la\,\phi_{m}(\la x)
\end{align}
and let $\la\to0$. The constants $N_\la$ have to be chosen in such a way that
the vacuum expectation values of these
rescaled fields do not scale to zero or diverge, but approach a finite limit.

The effect of the space-time rescaling $x \mapsto \la x$ is easily calculated:
Smearing the scaled field with a test
function $f\in\Ss(\Rl^2)$ amounts to evaluating $\phi_{m}$ on a scaled
testfunction $f_\la$,
\begin{align}
 f_\la(x)
 :=
 \la^{-2}f(\la^{-1}x)
 \,,\qquad
 \la>0\,,\;\,x\in\Rl^2\,.
\end{align}
The mass shell restrictions \eqref{eq:fm} of the Fourier transforms of such
scaled functions are given by scaling the
mass and the momentum,
\begin{align}\label{scale-f}
 f_\la^{m\pm}(p)
 =
 f^{\la m \pm}(\la p)
 \,.
\end{align}

As in the analysis of the free field \cite{Buchholz:1998vu}, two different
choices for the multiplicative
renormalization $N_\la$ are possible, namely $N_\la=1$ and
$N_\la=|\ln\la|^{-1/2}$. The latter choice corresponds to an
anomalous scaling of the field $\phi_m(f)$ when $f^{0\pm}(0) \neq 0$, which in
turn is due to the infrared divergence of
the $n$-point functions of the field in the massless limit. As in the case of
free fields, it can be expected that it
gives rise to an abelian tensor factor in the scaling limit algebra, at least if
$S(0) = 1$. We will however not
investigate this possibility any further here.

Choosing therefore $N_\la=1$, we now consider the $n$-point functions of the
rescaled field
\begin{align}\label{def:Wn}
 \We_{m}^{n,\la}(f_1,...,f_n)
 :=
 \langle\Om_m,\,\phi_{m}(f_{1\la})\cdots\phi_{m}(f_{n\la})\Om_m\rangle
 \,,
\end{align}
and study their limit as $\la\to0$.

\begin{proposition}
Let $m>0$, $S\in\SFlim$ and $f_1,...,f_n\in\Ss(\Rl^2)$ with $f_j^{0\pm}(0)=0$,
$j=1,...,n$. Then
 \begin{align}\label{wnlimit}
  \lim_{\la\to0}
  \We_{m}^{n,\la}(f_1,...,f_n)
  =
  \We_{0}^{n,1}(f_1,...,f_n)
  \,.
 \end{align}
 An analogous statement holds for the expectation values of the fields
$\phi_m'$, $\phi_0'$.
\end{proposition}
\begin{proof}
Both fields, $\phi_{m}$ and $\phi_{0}$, are defined as sums of certain creation
and annihilation operators, which change
the particle number by $\pm1$. Hence vacuum expectation values of products of an
odd number of field operators vanish,
i.e., the statement holds trivially if $n$ is odd. We may therefore assume that
$n=2k$ is even, and first consider the
vacuum expectation value of a particularly ordered product of creation and
annihilation operators. Using \eqref{eq:ZM},
\eqref{eq:ZdM} and \eqref{def:Pn}, we compute
 \begin{align}
   \langle
   &
   \Om_m,\,
 z_{m}(f_{1,\la}^{m-})\cdots z_{m}(f_{k,\la}^{m-})
 \zd_{m}(f_{k+1,\la}^{m+})\cdots \zd_{m}(f_{n,\la}^{m+})
 \Om_m\rangle
 \nonumber
 \\
 &=
 \sum_{\pi\in\frS_k}
 \langle \overline{f_{k,\la}^{m-}} \ot ... \ot\overline{f_{1,\la}^{m-}}
 \,,\,
 D_k(\pi)\,(f_{k+1,\la}^{m+}\ot...\ot f_{n,\la}^{m+})
 \rangle
 \nonumber
 \\
 &=
 \sum_{\pi\in\frS_k}
 \int \frac{d p_1}{\om_{1}^{m}}\cdots \frac{d p_k}{\om_{k}^{m}}
 \prod_{j=1}^k \Big(f_{k-j+1}^{\la m-}(\la p_j) f_{k+j}^{\la m+}(\la
p_{\pi(j)})\Big)
 \prod_{\substack{1\leq l<r\leq k \\  \pi(l)>\pi(r) }}
 	S_m(p_{\pi(l)}, p_{\pi(r)})
 \,,\label{eq:ScalarProd}
\end{align}
where we used the scaling relation \eqref{scale-f} in the last line.  Taking
into account the scaling relation
\eqref{eq:SM-scale} for $S_m$ and $dp/\om^m_p=d(\la p)/\om^{\la m}_{\la p}$, the
change of variables $p_j\to\la\,p_j$
yields
\begin{align}
 \langle&
 \Om_m,\,
 z_{m}(f_{1,\la}^{m-})\cdots z_{m}(f_{k,\la}^{m-})
 \zd_{m}(f_{k+1,\la}^{m+})\cdots \zd_{m}(f_{n,\la}^{m+})
 \Om_m\rangle
 \nonumber
 \\
 &=
 \sum_{\pi\in\frS_k}
 \int \frac{d p_1}{\om_{1}^{\la m}}\cdots \frac{d p_k}{\om_{k}^{\la m}}
 \prod_{j=1}^k \Big(f_{k-j+1}^{\la m-}(p_j) f_{k+j}^{\la m+}(p_{\pi(j)})\Big)
 \prod_{\substack{ 1\leq l<r\leq k \\ \pi(l)>\pi(r) }}
 	S_{\la m}(p_{\pi(l)}, p_{\pi(r)})
 \,.
 \label{zzdint}
 \end{align}
 In the limit $\la\to0$, the integrand converges pointwise to the corresponding
expression with $m=0$, which is
integrable because of our assumption on the test functions $f_j^{0\pm}$. For the
Schwartz class functions $f_j^{\la
m\pm}$, there exist $\la$-independent integrable bounds, and since $S_{\la m}$
has constant modulus 1, and $|\om_j^{\la
m}|^{-1}\leq|p_j|^{-1}$, we can use dominated convergence to conclude
 \begin{align}\label{wnlimit2}
  \lim_{\la\to0}
  \langle\Om_m,\,
 &
 z_{m}(f_{1,\la}^{m-})
 \cdots z_{m}(f_{k,\la}^{m-})
 \zd_{m}(f_{k+1,\la}^{m+})\cdots \zd_{m}(f_{n,\la}^{m+})
 \Om_m\rangle
 \\
 & =
 \sum_{\pi\in\frS_k}
 \int \frac{d p_1}{|p_1|}\cdots \frac{d p_k}{|p_k|}
 \prod_{j=1}^k (f_{k-j+1}^{0-}(p_j) f_{k+j}^{0+}(p_{\pi(j)}))
 \prod_{\substack{ 1\leq l<r\leq k \\ \pi(l)>\pi(r)} }
 	S_0(p_{\pi(l)},p_{\pi(r)})
 \nonumber
 \\
 &=
 \langle\Om_0,\,
  z_{0}(f_{1}^{0-})\cdots z_{0}(f_{k}^{0-})
 \zd_{0}(f_{k+1}^{0+})\cdots \zd_{0}(f_{n}^{0+})
 \Om_0\rangle
 \,.
 \nonumber
 \end{align}

 After this preparation, we consider the $(2k)$-point function of the field
$\phi_{m}$, and expand the fields into
creation and annihilation operators,
\begin{align*}
 \We_{m}^{2k,\la}(f_1,...,f_{2k})
 &=
 \langle\Om_m,\,
 \big(z_{m}(f_{1,\la}^{m-}) + \zd_{m}(f_{1,\la}^{m+})\big)
 \cdots
 \big(z_{m}(f_{2k,\la}^{m-}) + \zd_{m}(f_{2k,\la}^{m+})\big)
 \Om_m\rangle
 \,,
\end{align*}
and analogously for $m=0$. These are sums of $2^{2k}$ terms, each of which is
the vacuum expectation value of a
$(2k)$-fold product of $z_{m}$'s and $\zd_{m}$'s (respectively, $z_0$'s and
$\zd_0$'s). Because of the
annihilation/creation properties of these operators, all terms in which the
number of $z$'s is different from the number
of $\zd$'s vanish. So each non-zero term is of the form considered before, up to
a reshuffling of creation and
annihilation operators.

Picking any one of these terms, we can use the exchange relations of
Zamolodchikov's algebra to write the product of
creation and annihilation operators as a sum of products of the particular form
considered above, where all creation
operators stand to the right of all annihilation operators. The only difference
to the previous integral expressions is
that the reordering may reduce the number of integrations in \eqref{zzdint} --
due to the term $\om^m_p \delta(p-q)$ in
the Zamolodchikov's relation -- and introduce various factors of $S_{\la m}(p_a,
p_b)$, $a,b \in\{1,...,k\}$, in the
integrand as well as a permutation of the momenta $p_1,...,p_{k}$.

But the reorderings are the same for the case $m>0$ and $m=0$, and the
additional factors $S_{\la m}(p_a,p_b)$  converge
pointwise and uniformly bounded to their counterparts with $m=0$ in the limit
$\la\to0$. Thus the analogue of the limit
\eqref{wnlimit2} holds for an arbitrarily ordered product of $z$'s and $\zd$'s,
and \eqref{wnlimit} follows.
\end{proof}

According to this result, the scaling limit of $n$-point functions of the field
$\phi_m$ is given by the $n$-point
functions of the field $\phi_0$, with the same scattering function $S$. We will
then regard the $(0,S)$-model as the
short distance scaling limit of the $(m,S)$-model. For a discussion of the
relations with the
Buchholz-Verch scaling limit, we refer the reader to the  conclusions in
Sec.~\ref{section:conclusion}.

The massless wedge-localized fields obtained in the above limit are actually
{\em chiral}, and split into sums of
half\/line-localized fields on the two light rays. In the following, we will
indicate this split on a formal level; the
chiral fields will be defined more precisely in
Sec.~\ref{section:ChiralModels}. The localization regions of the
various fields appearing in this split can be visualized as in the picture below
(page \pageref{figure}). We will use
indices $\ri$/$\lef$ to distinguish between the right/left moving component
fields. (To avoid confusion, we notice
explicitly that this means that, e.g., the \emph{right} moving field is a
function of $x_\ri =  x_0-x_1$ only, and
therefore lives on the \emph{left} light ray, defined by $x_0+x_1= 0$.) Similar
to the fields $\phi_m'$, $\phi_m$ on
$\Rl^2$, operators with/without prime are localized on one or the other side of
a fixed light ray.

\begin{center}
	\includegraphics[width=8.5cm]{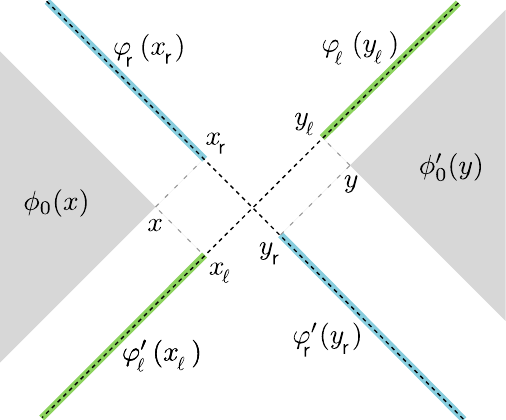}
	\\
	{\em Massless wedge- and halfline-localized quantum fields with their
localization regions}
	\label{figure}
\end{center}

It follows from \eqref{eq:fm}, \eqref{eq:phi} that the field $\phi_0$ is
formally defined by the operator valued
distribution
\begin{equation*}
\phi_0(x) = \int_\Rl \frac{dp}{2\pi
|p|}\big(e^{i(|p|x_0-px_1)}\zd_0(p)+e^{-i(|p|x_0-px_1)}z_0(p)\big).
\end{equation*}
Splitting now the integration in the sum of an integration over $(-\infty, 0)$
and one over $(0,+\infty)$, and
changing variable $p \to -p$ in the former integration, one gets
\begin{equation*}
\phi_0(x) = \frac{1}{\sqrt{2\pi}}(\varphi_\ri(x_\ri) + \varphi'_\lef(x_\lef))\,.
\end{equation*}
Here $x_\lef := x_0 + x_1$, $x_\ri := x_0-x_1$ are the left and right light ray
components of $x = (x_0,x_1)$, and
\begin{align}
\varphi_\ri(x_\ri) &:=
\int_0^{+\infty}\frac{dp}{\sqrt{2\pi}\,p}\big(e^{ipx_\ri}\zd_0(p)+e^{-ipx_\ri}
z_0(p)\big),
\label{eq:varphip}\\
\varphi'_\lef(x_\lef) &:=
\int_0^{+\infty}\frac{dp}{\sqrt{2\pi}\,p}\big(e^{ipx_\lef}\zd_0(-p)+e^{-ipx_\lef
}z_0(-p)\big),
\label{eq:varphim}
\end{align}
are two chiral (one-dimensional) fields living on the left/right light ray of
two-dimen\-sio\-nal Minkowski space. (In
order to avoid the infrared divergence which is apparent in the above integrals,
one
should actually consider the derivatives of these fields. At this formal level
this is not really relevant, but we will
consistently do so in the following section.) Notice also that, according
to~\eqref{eq:S0} and
\eqref{eq:ZFzz}--\eqref{eq:ZFzdz}, one has,
for $p, q > 0$,
\begin{align*}
z_0(-p)z_0(q) &= S(\infty) z_0(q)z_0(-p),\\
z_0(-p)\zd_0(q) &= S(\infty) \zd_0(q)z_0(-p).
\end{align*}
This implies that $\varphi'_\lef$ and $\varphi_\ri$ commute, resp.\ anticommute,
if $S(\infty) = +1$, resp.\ $S(\infty)
= -1$. Proceeding in the same way for the right-wedge field $\phi'_0$, one gets
an analogous split into two chiral
fields $\varphi'_\ri$, $\varphi_\lef$ defined by substituting $z^\#_0(p)$ with
$U_0(j)z^\#_0(p) U_0(j)^*$ in
formulas~\eqref{eq:varphip}, \eqref{eq:varphim}. It is then not difficult to
see, following the arguments
in~\cite{Lechner:2003}, that
\begin{equation*}
[\varphi_\lef(x_\lef), \varphi'_\lef(y_\lef)] = 0 \quad \text{if }x_\lef >
y_\lef
\end{equation*}
(see also Proposition~\ref{proposition:ChiralFields}.~\ref{it:PhiHalfLocal}
below). This shows that $\varphi_\lef$ and
$\varphi'_\lef$ can be interpreted as being localized in the right and
left half-line, respectively. An analogous statement holds
of course for $\varphi_\ri$, $\varphi'_\ri$.

The above formal manipulations suggest that the $(0,S)$-model can be written as
the (twisted, if $S(\infty) = -1$)
tensor product of two chiral models which are again defined in terms of the
Zamolodchikov-Faddeev
algebra~\eqref{eq:ZFzz}--\eqref{eq:ZFzdz} with $m = 0$ and $p, q > 0$. These
models will be rigorously defined in the
next section, and in Sec.~\ref{subsection:Splitting} we will show that such a
tensor product decomposition actually
holds.

\section{Chiral integrable models}\label{section:ChiralModels}

We saw in the previous section that the wedge-local fields generating the
massless $(0,S)$-models factorize into chiral
components. To analyze this connection in detail, it turns out to be most
convenient to first introduce the chiral
fields independently of the previously discussed models on two-dimensional
Minkowski space, and discuss the relation to
the $(0,S)$-models afterwards.

In this section, we will therefore be concerned with quantum fields on the real
line. The development of these models is
largely parallel to Sec.~\ref{subsection:2dmodels}, but has some distinctive
differences. Our construction will yield
dilation and translation covariant quantum field theory models on $\Rl$ (thought
of as either the right or left light
ray), with algebras of observables localized in half lines and intervals.

An important question is whether these models extend to \emph{conformally}
covariant theories on the circle. Using
results of \cite{GLW:extensions}, it turns out that this question is closely
related to the size of local algebras
associated with bounded intervals. This point will be discussed in detail in
Sec.~\ref{section:Conformal}.

\subsection[S-symmetric Fock space and Zamolodchikov's algebra]{$S$-symmetric
Fock space and Zamolodchikov's algebra on
the light ray}\label{subsection:ChiralZamolodchikov}

As before, we start from a scattering function $S \in \SFlim$. We first define
the Hilbert space of the theory. Our
``chiral'' single particle space is given by
\begin{equation*}
\Hil_{1} := L^2(\Rl,d\beta)\,.
\end{equation*}
The variable $\beta$ is meant to be related to the momentum $p$ by $p =
e^\beta$, as will become clear in
\eqref{eq:ChiralSymmetry} below. Like in \eqref{def:Pn}, we have a unitary
action $D_n$ of the permutation group
$\frS_n$ on $\Hil_{1}\tp{n}=L^2(\Rl^n,d\bbeta)$ which acts on transpositions
$\tau_k$ by
\begin{equation*}
(D_n(\tau_k)\Psi_n)(\beta_1,....,\beta_n)
=
S(\beta_{k+1}-\beta_k)\cdot \Psi_n(\beta_1,...,\beta_{k+1},\beta_k,...,\beta_n)
\,,\quad
\Psi_n\in L^2(\Rl^n,d\bbeta)\,.
\end{equation*}
Again, our $n$-particle space $\Hil_{n}$ is defined as the $D_n$-invariant
subspace of $\Hil_{1}\tp{n}$, and the
projector onto it is denoted as $P_n$. We define the Fock space
$\Hil:=\bigoplus_{n \geq 0} \Hil_{n}$ and its subspace
$\DD \subset \Hil $ consisting of vectors of bounded particle number, i.e., of
terminating sequences.

We proceed to a representation of the space-time symmetries. On $\Hil$, we
consider a unitary representation $U$ of the
affine group $G$ of $\Rl$, consisting of translations and dilations, $\Rl\ni
\xi\mapsto e^{\la}\xi+\xi'$, and the reflection, $j(\xi):=-\xi$. For translations and
dilations, it is defined as, $\xi,\la\in\Rl$,
\begin{equation}\label{eq:ChiralSymmetry}
(U(\xi,\la)\Psi)_n(\beta_1,...,\beta_n)
:=
e^{i\xi(e^{\beta_1}+...+e^{\beta_n})}\cdot \Psi_n(\beta_1+\la,...,\beta_n+\la)\,,
\end{equation}
and the reflection $j$ is represented antiunitarily by
\begin{equation} \label{eq:ChiralReflection}
(U(j)\Psi)_n(\beta_1,...,\beta_n)
:=
\overline{\Psi_n(\beta_n,...,\beta_1)}
=
\prod_{1\leq l<r\leq n}
S(\beta_r-\beta_l)\cdot \overline{\Psi_n(\beta_1,...,\beta_n)}
.
\end{equation}
Compare this with the two-dimensional case in \eqref{eq:2dTranslationDilation},
\eqref{eq:Uj}.
We will also use the shorthand notation $U(\xi):=U(\xi,0)$ for pure translations,
and note here that this one parameter
group has a positive generator, $H$. Up to scalar multiples,
$\Om:=1\oplus0\oplus 0 ...$ is the only $U$-invariant
vector in $\Hil$; it will play the role of the vacuum vector.

As in \eqref{eq:ZM}, we will make use of ``$S$-symmetrized'' annihilation and
creation operators, which we label $y$ and
$\yd$, in order to distinguish them from $z_m,\zd_m$, since they will take
rapidities rather than momenta as arguments.
For $\psi\in\Hil_1$, $\Phi\in\DD$, they act by
\begin{align}\label{eq:def-yyd}
 (\yd(\psi)\Phi)_n := \sqrt{n}P_n(\psi\otimes\Phi_{n-1})
\,,\qquad
y(\psi)
:=
\yd(\overline{\psi})^*\,.
\end{align}
Except for the special case $S=-1$, these are unbounded operators containing
$\DD$ in their domains. Under symmetry
transformations, they behave like
\begin{align}\label{eq:yyd-covariance}
 U(\xi,\la) \yd(\psi)U(\xi,\la)^{-1} &= \yd(U(\xi,\la)\psi)
\,,\\
 U(\xi,\la) y(\psi)U(\xi,\la)^{-1} &= y(U(-\xi,\la)\psi)
\,,
\end{align}
whereas with respect to the reflection $j$, no such transformation formula
holds.

From time to time, we will also work with operator-valued distributions
$y(\beta),\yd(\beta)$, $\beta\in\Rl$, related to
the above operators by the formal integrals $y^\#(\psi)=\int
d\beta\,\psi(\beta)y^\#(\beta)$. They satisfy the relations
of the Zamolodchikov-Faddeev algebra in the form
\begin{align}
\label{eq:zfy-1}
 y(\beta_1)y(\beta_2) &= S(\beta_1-\beta_2)\, y(\beta_2)y(\beta_1)\,,
\\
\label{eq:zfy-2}
 \yd(\beta_1)\yd(\beta_2) &= S(\beta_1-\beta_2)\, \yd(\beta_2)\yd(\beta_1)\,,
\\
\label{eq:zfy-3}
y(\beta_1)\yd(\beta_2) &=
S(\beta_2-\beta_1)\,\yd(\beta_2)y(\beta_1)+\delta(\beta_1-\beta_2)\cdot 1\,.
\end{align}
It is interesting to note that these are exactly the same relations as used in
\emph{massive} two-dimensional models,
written in terms of rapidities \cite{Lechner:2008}. We will however see that the
interpretation in terms of wedge-local
observables must be modified in the chiral case.

\subsection{Half-local quantum fields and observable
algebras}\label{subsection:ChiralFields}

We now set out to construct a pair of quantum fields on $\Hil$ as sums of
Zamolodchikov type creation and annihilation
operators, analogous to the two-dimensional case in
Sec.~\ref{subsection:2dmodels}. For the one-dimensional case, these
quantum fields will be localized in half-lines, rather than in wedge regions.
While we employ largely the same ideas as
in the massive two-dimensional case \cite{Lechner:2008}, the chiral situation
makes some modifications necessary, so
that we will need to look into the construction in more detail.

We first introduce the necessary test functions and discuss their properties.
For $f\in\Ss(\Rl), \varphi\in C_0^\infty(\Rl)$, we define their Fourier
transforms and the positive/negative frequency
components of those with the following conventions.
\begin{align}\label{eq:HatTransform}
\fhat^\pm(\beta)
&:=
\pm i\,e^\beta\,\fti(\pm e^\beta) = \pm \frac{i\,e^\beta}{\sqrt{2 \pi}} \int
f(\xi) \exp(\pm i e^\beta \xi) d\xi
\,,
\\
\label{eq:CheckTransform}
\check{\varphi}^\pm(\xi)
&:=
\mp\frac{i}{\sqrt{2\pi}}\int d\beta\,\varphi(\beta)\,e^{\mp i\xi e^\beta}
=
\mp\frac{i}{\sqrt{2\pi}}\int_0^\infty dp\,\frac{\varphi(\log p)}{p}\,e^{\mp ip\xi}
\,.
\end{align}

\begin{lemma}\label{lemma:fpm}
Let $f\in\Ss(\Rl), \varphi\in C_0^\infty(\Rl)$.
\begin{enumerate}
 \item $\fhat^\pm, \check{\varphi}^\pm\in\Ss(\Rl)$. As maps from $\Ss(\Rl)$ to
$L^2(\Rl)$, $f\mapsto\fhat^\pm$ are
continuous.
 \item For $\varphi\in C_0^\infty(\Rl)$, there holds
\begin{align*}
(\check{\varphi}^\pm\hat{)}^\pm = \varphi
\,,\qquad
(\check{\varphi}^\pm\hat{)}^\mp = 0
\,.
\end{align*}
\item Let $f^{\xi,\la}(\xi'):=f(e^{-\la}(\xi'-\xi))$ and $f^j(\xi):=\overline{f(-\xi)}$. Then
\begin{align}\label{eq:fpm-covariance}
 (\widehat{f^{\xi,\la}})^\pm(\beta) = e^{\pm i\xi e^\beta}\,\fhat^\pm(\beta+\la)
\,,\qquad
 (\widehat{f^j})^\pm(\beta) = -\overline{\fhat^\pm(\beta)}
\,,\qquad
\widehat{\fbar}^\pm = \overline{\fhat^\mp}\,.
\end{align}
\item Let $f,g\in\Ss(\Rl)$, with $\supp f\subset \Rl_+$, $\supp g\subset\Rl_-$.
Then $\fhat^+$ and $\ghat^-$ have
bounded analytic extensions to the strip $\Strip(0,\pi)$, and
$|\fhat^+(\beta+i\la)|,|\ghat^-(\beta+i\la)|\to0$ as
$\beta\to\pm\infty$, uniformly in $\la\in[0,\pi]$. The boundary values at
$\operatorname{Im}\beta=\pi$ are
\begin{align}\label{eq:fpm-boundary}
 \fhat^+(\beta+i\pi) = \fhat^-(\beta)\,,\qquad \ghat^-(\beta+i\pi) =
\ghat^+(\beta)\,,\qquad \beta\in\Rl\,.
\end{align}
If $\supp f\subset (r,\infty)$ and $\supp g\subset(-\infty,-r)$ for some $r>0$,
then there exist $c,c'>0$ such that
\begin{align}\label{eq:fpm-bnd-1}
|\fhat^+(\beta+i\la)| \leq c\,e^{-re^\beta \sin\la}
\,,\quad
|\ghat^-(\beta+i\la)| \leq c'\,e^{-re^\beta \sin\la}
\,,\qquad 0\leq\la\leq\pi\,.
\end{align}
\end{enumerate}
\end{lemma}
\begin{proof}
a) It is clear that $\fhat^\pm\in\Ss(\Rl)$, and by considering the second
formula in \eqref{eq:CheckTransform},
$\check{\varphi}^\pm$ is seen to be the Fourier transform of a function in
$C_0^\infty(\Rl)$, and hence of Schwartz
class, too. Since $\fti\in\Ss(\Rl)$, one gets the bound $|\fhat^\pm(\beta)|\leq
c_\pm(f)\cdot e^{-|\beta|}$ for some
Schwartz seminorm $c_\pm(f)$, which implies the claimed continuity by estimating
the $L^2$-norm of $\fhat^\pm$.

b) By its definition \eqref{eq:CheckTransform}, $\check \varphi^{\pm}$ is the
inverse Fourier transform of the function
$p \mapsto \mp i \theta(\pm p) \varphi(\log \lvert p \rvert )/\lvert p \rvert$,
where $\theta$ denotes the step
function. The statement now follows from the Fourier inversion formula.

c) This is obtained by straightforward calculation.

d) The analyticity of $\fhat^+$ in $\Strip(0,\pi)$ follows from the analyticity
of $\fti$ in the upper complex half
plane (since $\supp f\subset\Rl_+$), and the fact that the exponential function
maps $\Strip(0,\pi)$ onto the upper half
plane. The uniform bound follows from the estimate
\begin{equation}
\begin{aligned}
|\fhat^+(\beta+i\la)|
&=
\frac{1}{\sqrt{2\pi}}
\left|\int_0^\infty d\xi\,\partial_\xi f(\xi) e^{i\xi e^{\beta+i\la}}\right|
\\
&\leq
\frac{1}{\sqrt{2\pi}}
\int_0^\infty d\xi\,|\partial_\xi f(\xi)| e^{-\xi e^\beta\sin\la}
\leq
\frac{\|\partial_\xi f \|_1}{\sqrt{2\pi}}
\,,
\end{aligned}
\end{equation}
where in the last step we used $\xi>0$, $0\leq\la\leq\pi$.

The claimed boundary value follows directly from the definition of $\fhat^+$ in
\eqref{eq:HatTransform}:
\begin{align}
\fhat^+(\beta+i\pi)
=
i\,e^{\beta+i\pi}\,\fti(e^{\beta+i\pi})
=
- i\,e^\beta\,\fti(-e^\beta)
=
\fhat^-(\beta)
\,.
\end{align}
So $\fhat^+(\beta)$ and $\fhat^+(\beta+i\pi)$ converge to zero for
$\beta\to\pm\infty$. Since these functions are
bounded and analytic in $\Strip(0,\pi)$, it follows that also
$|\fhat^+(\beta+i\la)|\to0$ as $\beta\to\pm\infty$,
uniformly in $\la\in[0,\pi]$ -- see, for example, \cite[Cor.~1.4.5]{Boas:1954}.

To obtain the sharpened bound \eqref{eq:fpm-bnd-1}, note that if $\supp
f\subset(r,\infty)$, then $f^{-r,0}$ (cf.~part
c)) has support in $\Rl_+$, and
$\widehat{f^{-r,0}}^+(\beta)=e^{-ire^\beta}\fhat^+(\beta)$ due to
\eqref{eq:fpm-covariance}. So there exists $c>0$ such that for any
$\la\in[0,\pi]$,
\begin{equation}
c
>
|e^{-ire^{\beta+i\la}}\fhat^+(\beta+i\la)|
=
e^{re^\beta\sin\la}|\fhat^+(\beta+i\la)|\,,
\end{equation}
which implies \eqref{eq:fpm-bnd-1}.

Finally, given $g\in\Ss(\Rl)$ with $\supp g\subset \Rl_-$ respectively $\supp
g\subset(-\infty,-r)$, all corresponding
statements about $\ghat^-$ follow from the previous arguments by considering
$f(\xi):=g(-\xi)$, since $\supp f\subset\Rl_+$,
and $\fhat^+=-\ghat^-$.
\end{proof}

After these preparations, we define for $f\in\Ss(\Rl)$ the two field operators,
\begin{align}\label{eq:ChiralFieldDef}
  \phi(f) &:= \yd(\fhat^+) + y(\fhat^-)\,,
  \\ \label{eq:ChiralConjugateFieldDef}
  \phi'(f) &:= U(j)\phi(f^j)U(j)
  \,.
\end{align}

These fields should be thought of as the derivatives of the left/right chiral
fields $\varphi_{l/r}^{[\prime]}$
appearing in the decomposition of the massless two-dimensional field $\phi_0$,
cf. also the figure on page
\pageref{figure}.

For reference, we note the ``unsmeared'', distributional version of
\eqref{eq:ChiralFieldDef}:
\begin{equation}\label{eq:ChiralUnsmearedField}
  \phi(\xi) = i \int \frac{d\beta}{\sqrt{2\pi}} \, e^\beta \Big(
  e^{ie^\beta \xi} \yd(\beta)
 - e^{-ie^\beta \xi} y(\beta)
\Big).
\end{equation}
The main features of these fields can largely be obtained in the same way as in
\cite{Lechner:2003}.
\begin{proposition}\label{proposition:ChiralFields}
$\phi$ and $\phi'$ have the following properties.
\begin{enumerate}
 \item \label{it:PhiTempered}
The map $f\mapsto\phi(f)$ is an operator-valued tempered distribution such that
$\DD$ is contained in the domain of
$\phi(f)$ for all $f\in\Ss(\Rl)$. For real $f$, the operator $\phi(f)$ is
essentially self-adjoint, with elements from
$\DD$ as entire analytic vectors.
\item \label{it:PhiCovariant}
$\phi$ transforms covariantly under the representation $U$ of the connected
component of the affine group, i.e.,
\begin{equation}\label{phi-covariance}
 U(\xi,\la)\phi(f)U(\xi,\la)^{-1}=\phi(f^{\xi,\la})\,.
\end{equation}
\item \label{it:ReehSchlieder}
The Reeh-Schlieder property holds, i.e., for any non-empty open interval
$I\subset\Rl$, the set
\begin{equation}
\operatorname{span}\{\phi(f_1)\cdots\phi(f_n)\Om\,:\,f_1,...,f_n\in\Ss(I),
\;n\in\Nl_0\}
\end{equation}
is dense in $\Hil$.
\item \label{it:PhiHalfLocal}
$\phi$ and $\phi'$ are relatively half-local in the following sense: If
$f,g\in\Ss(\Rl)$ satisfy $\supp f\subset
(a,\infty)$, $\supp g\subset (-\infty,a)$ for some $a\in\Rl$, then
\begin{align}
 [\phi(f),\,\phi'(g)]\Psi=0\,\qquad\text{for all }\Psi\in\DD\,.
\end{align}
\end{enumerate}
Statements \ref{it:PhiTempered}-\ref{it:ReehSchlieder} also hold when $\phi$ is
replaced with $\phi'$.
\end{proposition}
\begin{proof}
a) It is clear from the definition of $\phi(f)$ that these operators always
contain $\DD$ in their domains and depend
complex linearly on $f\in\Ss(\Rl)$. Taking into account that the restrictions of
the creation/annihilation operators to
an $n$-particle space $\Hil_{n}$ are bounded, $\|y^\#(\psi)\lceil\Hil_{n}\|\leq
\sqrt{n+1} \|\psi\|_{\Hil_{1}}$, and the
continuity of $\Ss(\Rl)\ni f\mapsto\fhat^\pm\in\Hil_{1}$ established in Lemma
\ref{lemma:fpm} a), it follows that $\phi$
is an operator-valued tempered distribution.

In view of \eqref{eq:def-yyd} and \eqref{eq:fpm-covariance}, we have
\begin{align*}
\phi(f)^*
=
\big(\yd(\fhat^+)+y(\fhat^-)\big)^*
\supset
y(\overline{\fhat^+})+\yd(\overline{\fhat^-})
=
y(\widehat{\fbar}^-)+\yd(\widehat{\fbar}^+)
=
\phi(\fbar)
\,.
\end{align*}
This shows that $\phi(f)$ is hermitian for real $f$, and the proof of essential
self-adjointness can now be completed as
in \cite[Prop.~1]{Lechner:2003} by showing that any vector in $\DD$ is entire
analytic for $\phi(f)$.

b) This is a direct consequence of \eqref{eq:yyd-covariance} and
\eqref{eq:fpm-covariance}.

\ref{it:ReehSchlieder} Let $\Pol(I)$ denote the algebra generated by all
polynomials in the field $\phi(f)$ with $\supp
f \subset I$. By standard analyticity arguments making use of the positivity of
the generator of $\xi\mapsto U(\xi)$, it
follows that $\Pol(I)\Om$ is dense in $\Hil$ if and only if $\Pol(\Rl)\Om$ is
dense in $\Hil$. But given any $\varphi\in
C_0^\infty(\Rl)$, the function $f:=\check{\varphi}^+\in\Ss(\Rl)$ satisfies
$\fhat^+=\varphi$ and $\fhat^-=0$ (Lemma
\ref{lemma:fpm} b), and hence $\yd(\varphi)=\phi(f)\in\Pol(\Rl)$. Since
$C_0^\infty(\Rl)$ is dense in $\Hil_1$,
polynomials in the $\yd(\varphi)$ create a dense set from $\Om$.

The proofs of statements a)--c) for the field $\phi'$ are completely analogous.

d) Since the Zamolodchikov-Faddeev relations \eqref{eq:zfy-1}--\eqref{eq:zfy-3}
are the same as in the massive case in
rapidity space, we can establish the following commutation relations in complete
analogy to \cite[Lemma
4]{Lechner:2003}:
\begin{equation} \label{eq:yypCommutator}
\begin{aligned}
 \lbrack y(\psi_1),\,U(j)y(\psi_2)U(j) \rbrack
&=0
\,,\\
 [\yd(\psi_1),\,U(j)\yd(\psi_2)U(j)]
&=0
,\\
([U(j)y(\overline{\psi_1})U(j),\,\yd(\psi_2)]\Phi)_n(\bbeta)
&=
C^{\psi_1,\psi_2,+}_n(\bbeta)\cdot\Phi_n(\bbeta),
\\
([U(j)\yd(\overline{\psi_1})U(j),\,y(\psi_2)]\Phi)_n(\bbeta)
&=
C^{\psi_1,\psi_2,-}_n(\bbeta)\cdot\Phi_n(\bbeta),
\end{aligned}
\end{equation}
where $\psi_1,\psi_2\in\Hil_{1}$, $\Phi\in\DD$, and
\begin{equation}\label{eq:commutor-function}
C^{\psi_1,\psi_2,\pm}_n(\bbeta)
=
\pm\int d\beta_0\,\psi_1(\beta_0)\psi_2(\beta_0)\prod_{k=1}^n
S(\pm\beta_0\mp\beta_k)
\,.
\end{equation}
In view of the definition of the fields $\phi$ and $\phi'$, the commutator
takes the form
\begin{equation}
\begin{aligned}
\lbrack \phi(f),\phi'(g) \rbrack \Psi_n
&=
-[\yd(\fhat^+)+y(\fhat^-),\,U(j)\yd(\overline{\ghat^+})U(j)+U(j)y(\overline{
\ghat^-})U(j)]\Psi_n
\\
&=
(C_n^{\fhat^+,\ghat^-,+}+C_n^{\fhat^-,\ghat^+,-})\Psi_n
\, \quad \text{for } \Psi_n\in\Hil_{ n}.
\end{aligned}
\end{equation}
Due to the translational covariance of $\phi$ and $\phi'$, it is sufficient to
consider the case $a=0$, i.e., $\supp
f\subset\Rl_+$, $\supp g\subset\Rl_-$. To show that
$C_n^{\fhat^+,\ghat^-,+}+C_n^{\fhat^-,\ghat^+,-}=0$, we note that in
the integral
\begin{align}\label{eq:commutator-kernel}
C_n^{\fhat^+,\ghat^-,+}(\bbeta)
&=
\int d\beta_0\,\fhat^+(\beta_0)\ghat^-(\beta_0)\prod_{k=1}^n S(\beta_0-\beta_k)
\,,
\end{align}
all three functions, $\fhat^+,\ghat^-$, and $\beta_0\mapsto S(\beta_0-\beta_k)$,
have analytic continuations to the
strip $\Strip(0,\pi)$ (Def.~\ref{definition:ScatteringFunctions} and
Lemma~\ref{lemma:fpm} d)). According to Definition
\ref{definition:ScatteringFunctions}, the continuation of $S$ is bounded on this
strip, whereas according to Lemma
\ref{lemma:fpm} d), the functions $\fhat^+(\beta_0+i\la),\ghat^-(\beta_0+i\la)$
converge to zero for
$\beta_0\to\pm\infty$ uniformly in $\la\in[0,\pi]$. This implies that we can
shift the contour of integration from $\Rl$
to $\Rl+i\pi$ in \eqref{eq:commutator-kernel}. As the boundary values of the
integrated functions are given by
$\fhat^+(\beta_0+i\pi)=\fhat^-(\beta_0)$,
$\ghat^-(\beta_0+i\pi)=\ghat^+(\beta_0)$, and
$S(\beta_0+i\pi-\beta_k)=S(\beta_k-\beta_0)$, comparison with
\eqref{eq:commutor-function} shows
$C_n^{\fhat^+,\ghat^-,+}+C_n^{\fhat^-,\ghat^+,-}=0$.
\end{proof}

Proceeding to the algebraic formulation, we denote the self-adjoint closures of
$\phi(f)$ and $\phi'(f)$ (with $f$ real)
by the same symbols, and introduce the von Neumann algebras generated by them,
\begin{align}\label{eq:def-M}
 \M &:= \{e^{i\phi(f)}\,:\,f\in\Ss(\Rl_+)\;\text{real}\}''
\,,\\
\Mhat &:= \{e^{i\phi'(f)}\,:\,f\in\Ss(\Rl_-)\;\text{real}\}''
\,.
\end{align}
\pagebreak
\begin{theorem}\label{Thm:M}
The algebras $\M$ and $\Mhat$ have the following properties.
 \begin{enumerate}
  \item For $\xi\geq0, \la\in\Rl$, we have
\begin{align}\label{M-covariance}
 U(\xi,\la)\M U(\xi,\la)^{-1} \subset \M\,.
\end{align}
 \item The vector $\Om$ is cyclic and separating for $\M$.
\item The Tomita-Takesaki modular data of $(\M,\Om)$ are
\begin{align}
\Delta^{it}=U(0,-2\pi t)
\,,\qquad J = U(j)\,.
\end{align}
\item \label{it:MDuality}
$\Mhat=\M'$.
 \end{enumerate}
\end{theorem}
\begin{proof}
a) Given $f\in\Ss(\Rl_+)$ and $\xi\geq0$, $\la\in\Rl$, also $f^{\xi,\la}$  lies in
$\Ss(\Rl_+)$ by
\eqref{eq:fpm-covariance}. Since $U(\xi,\la)\M U(\xi,\la)^{-1}$ is generated by
$U(\xi,\la) e^{i\phi(f)}
U(\xi,\la)^{-1}=e^{i\phi(f^{\xi,\la})}$, cf.~\eqref{phi-covariance}, the claim
follows.

b) Taking into account that the field operator $\phi(f)$ is self-adjoint, we can
use standard arguments (see, for
example, \cite{BorYng:positivity}) to derive the cyclicity of $\Om$ for $\M$
from the cyclicity of $\Om$ for $\phi$,
which was established in Proposition \ref{proposition:ChiralFields} c).

Next we note that our fields $\phi,\phi'$ have the vacuum $\Om$ as an analytic
vector (Prop.
\ref{proposition:ChiralFields}~\ref{it:PhiTempered}), so that we can apply the
results of \cite{BorchersZimmermann:1963}
to conclude that also the unitaries $e^{i\phi(f)}, e^{i\phi'(g)}$ commute for
real $f,g$ with $\supp f\subset\Rl_+$,
$\supp g\subset\Rl_-$. That is, $\Mhat \subset\M'$. But $\Om$ is cyclic for
$\Mhat$ by the same argument as above, and
hence $\Om$ separates $\M$.

c) The proof of this claim works precisely as in
\cite[Prop.~3.1]{BuchholzLechner:2004} by exploiting the commutation
relations between $U(x)$ and $\Delta^{it}$, $J$, which are known from a theorem
of Borchers \cite{Borchers:1992}.

d) By definition of $\phi'$, we have $\Mhat=U(j)\M U(j)$. But $U(j)$ coincides
with the modular conjugation of
$(\M,\Om)$, and hence, by Tomita's theorem, $\Mhat=U(j)\M U(j)=J\M J=\M'$.
\end{proof}

\subsection{Local operators} \label{subsection:LocalChiral}

So far we have constructed a Hilbert space $\Hil$, a representation $U$ of $G$
on $\Hil$, and a von Neumann algebra
$\M\subset\B(\Hil)$ associated with a scattering function $S\in\SFlim$, such
that these data are compatible in the sense
of Theorem \ref{Thm:M}. Given such objects, we now recall how a corresponding
local field theory can be constructed. The
first step is to define a family of von Neumann algebras associated with
intervals, $-\infty<a<b<\infty$, as
\begin{equation}\label{eq:ChiralAlgebrasIntersect}
 \A(a,b) := U(a)\M U(a)^{-1} \cap U(b)\M' U(b)^{-1}\,.
\end{equation}
For general subsets $R$ of $\Rl$ we set
\begin{equation} \label{eq:AlgGenSubset}
 \A(R) := \bigvee_{(a,b)\subset R}\A(a,b)\,.
\end{equation}
This defines in particular the locally generated half line algebras
$\A(\Rl_+)\subset\M$ and $\A(\Rl_-)\subset\M'$, as
well as the global algebra $\A:=\A(\Rl)$.

The following properties of the assignment $I\mapsto\A(I)$ are all
straightforward consequences of Theorem \ref{Thm:M},
so that we can omit the proof.

\begin{proposition}
The map $I\mapsto\A(I)$ is an isotonous net of von Neumann algebras on $\Hil$
which transforms covariantly under the
affine group $G$,
\begin{align}
 U(g)\A(I)U(g)^{-1}=\A(gI)\,,\qquad g\in G\,.
\end{align}
This net of algebras is local in the sense that
\begin{align}
 \A(I_1)\subset\A(I_2)'
\qquad\text{whenever}\qquad I_1\cap I_2=\emptyset
\,.
\end{align}
\end{proposition}

Note that no statement regarding the size of the algebras $\A(I)$ is made here.
We shall see in
Sec.~\ref{section:Conformal} that this question is closely related to the
existence of conformal symmetry. However,
there is one restriction on the size of $\A(I)$ that we can compute directly: We
will show that all local operators
commute with $S(\infty)^N$, where $(N\Psi)_n:=n\cdot\Psi_n$ is the number
operator on $\Hil$; this limits the size of
$\A(I)$ in the case $S(\infty)=-1$. In the following preparatory lemma,
$P_n\in\B(\Hil)$  is understood as the
orthogonal projection onto $\Hil_n$.

\begin{lemma} \label{lemma:DilationConvergence}
Let $\psi_1,\psi_2 \in \Ss(\Rl)$, $n \in \Nl_0$.
The following sequences of bounded operators converge to zero in the weak
operator topology as $\lambda \to \infty$.
\begin{gather}
\label{eq:YLimit1}
  y(U(0,\lambda)\psi_1) \,  U(j)  \, y(U(0,\lambda)\psi_2) P_n
\\
\label{eq:YLimit2}
  \yd(U(0,\lambda)\psi_1) \,  U(j)  \, \yd(U(0,\lambda)\psi_2) P_n
\\
\label{eq:YLimit3}
\yd(U(0,\lambda)\psi_1) \,  U(j)  \, y(U(0,\lambda)\psi_2) P_n
\\
\label{eq:YLimit4}
  [ y(U(0,\lambda)\psi_1) , \, U(j) \yd(U(0,\lambda)\psi_2) U(j) ] -
\hrskp{\psi_2}{\psi_1} S(\infty)^N
\end{gather}
\end{lemma}
\begin{proof}
Let $\Psi_n \in \Hil_n \cap \Ss(\Rl^n)$. Expanding the definition
\eqref{eq:def-yyd} of the annihilation operator as in
\eqref{eq:ZM}, we find, $k=1,2$,
\begin{equation*}
 \| y( U(0,\lambda) \psi_k ) \Psi_n \|^2
=
n \int d^{n-1} \bbeta \, d\beta_0 \, d\beta_0' \,
  \psi_k(\beta_0 + \lambda)
 \overline{ \psi_k(\beta_0' + \lambda)  }
 \Psi_n(\beta_0,\bbeta)
 \overline{ \Psi_n(\beta_0',\bbeta) }.
\end{equation*}
As $\la\to\infty$, the integrand goes to zero pointwise, and since the functions
are all of Schwartz class, we can apply
the dominated convergence theorem to prove that
$ y( U(0,\lambda) \psi_k ) \Psi_n \to 0$ in Hilbert space norm.
On the other hand, we have the bound
$\|y(U(0,\lambda) \psi_k) P_n \| \leq \sqrt{n} \|\psi_k\|$, uniform in
$\lambda$.
Hence
\begin{equation}\label{eq:ySingleLimit}
\lim_{\la\to\infty}y( U(0,\lambda) \psi_k ) P_n = 0 \; \text{ in the strong
operator topology.}
\end{equation}
By another application of the uniform bound, and using $\|U(j)\|=1$, the
operator \eqref{eq:YLimit1}
converges to zero strongly.
The adjoint of this operator then vanishes in the weak operator topology. Since
this adjoint differs from
\eqref{eq:YLimit2} only by trivial redefinitions, the second claim follows. For
proving that \eqref{eq:YLimit3}
converges weakly to zero in the limit $\la\to \infty$, we just need to apply
\eqref{eq:ySingleLimit} on both sides of
the scalar product.

For the operator \eqref{eq:YLimit4}, we apply \eqref{eq:yypCommutator} to
obtain, with $\Phi_n,\Psi_n \in \Hil_n \cap
\Ss(\Rl^n)$,
\begin{multline}
 \hrskp{ \Phi_n} { \Big( [ y(U(0,\lambda) \psi_1) ,U(j) \yd(U(0,\lambda)\psi_2)
U(j) ]- \hrskp{\psi_2}{\psi_1 }
S(\infty)^N  \Big) \Psi_n }
 \\
=
  \int d^n \bbeta \int d\beta_0\,
   \overline{\Phi_n(\bbeta)} \Psi_n(\bbeta)
   \psi_1(\beta_0) \overline{\psi_2(\beta_0)}
  \Big(\prod_{l=1}^n S(\beta_l-\beta_0+\lambda) - S(\infty)^n\Big).
\end{multline}
The integrand tends to zero pointwise, and by another application of the
dominated convergence theorem, it follows that
the above matrix element vanishes as $\lambda \to \infty$. All matrix elements
of \eqref{eq:YLimit4} between vectors of
different particle number vanish identically. As \eqref{eq:YLimit4} is bounded
in operator norm, uniform in $\lambda$,
and $\Phi_n,\Psi_n$ were chosen from a total set, the operator
\eqref{eq:YLimit4} tends to zero in the weak operator
topology.
\end{proof}

As a consequence, all local operators are even with respect to the particle
number in the case $S(\infty)=-1$.

\begin{proposition} \label{proposition:EvenOps}
If $A \in \A(I)$ for some bounded interval $I$, then
$[A, S(\infty)^N] = 0$.
\end{proposition}
\begin{proof}
	Without loss of generality, let $I=(-1,1)$. We choose $g \in
\Ss(1,\infty)$ and $g' \in \Ss(-\infty,-1)$ fixed
such that $\hrskp{\overline{ \hat g^- }}{\hat g^{\prime +}}\neq 0$, and set
$f^{[\prime]} := g^{[\prime]0,\lambda}$ with
$\lambda \geq 0$. For any such $\la$, the (closed) field operator $\phi(f)$ is
affiliated with $U(1)\M U(1)^{-1}$, and
$\phi'(f')$ is affiliated with $U(-1)\M U(-1)^{-1}$. Since both
fields contain $\DD$ in their domains and leave this
subspace invariant, this implies that their product $\phi(f) \phi'(f')$ commutes
with
$\A(I)=U(1)\M'U(1)^{-1}\cap U(-1)\M U(-1)^{-1}$ on $\DD$, cf.
\eqref{eq:ChiralAlgebrasIntersect} and
Theorem~\ref{Thm:M}~\ref{it:MDuality}. Hence we find, $\Phi,\Psi \in \DD$,
	\begin{equation} \label{eq:afieldcomm}
		\hrskp{\Phi}{ [A,  \phi(f) \phi'(f') ]  \Psi} = 0
		\,.
	\end{equation}
	We can write
	\begin{align}
		\notag
		-\phi(f) \phi'(f')
		&=
		\Big(\yd(\hat f^+) + y(\hat f^-) \Big)
		U(j) \Big( \yd( \overline{ \hat f^{\prime +} }) + y( \overline{
\hat f^{\prime -} })  \Big) U(j)
		\\ \notag
		&\left.
		\begin{aligned}
			= &\hphantom{+}\yd(\hat f^+) U(j)\yd(\overline{\hat
f^{\prime +}})U(j)
			+ \yd(\hat f^+) U(j) y(\overline{\hat f^{\prime -}})
U(j)
			\\
			& +U(j) \yd(\overline{\hat f^{\prime +}}) U(j) y(\hat
f^-)
			+ y(\hat f^-) U(j) y(\overline{\hat f^{\prime -}}) U(j)
		\end{aligned}
		\right\} (*)
		\\
		& \hphantom{=}\;\,+ [y(\hat f^-) ,U(j) \yd(\overline{\hat
f^{\prime +}}) U(j)] .
	\end{align}
	Inserted into the matrix element \eqref{eq:afieldcomm}, the expression
$(\ast)$ vanishes as $\lambda \to \infty$
due to Lemma~\ref{lemma:DilationConvergence}. In the same way, the remaining
commutator converges to
$\hrskp{\overline{\hat g^-}}{\hat g^{\prime +}} S(\infty)^N$. Since
$\hrskp{\overline{\hat g^-}}{\hat g^{\prime +}} \neq
0$, the claim follows.
\end{proof}

Thus, at least for the class of models with $S(\infty)=-1$, we have some
restriction on the size of the local algebras
$\A(I)$; in particular, the inclusion $\A(0,\infty) \subset \M$ is proper in
these cases.

\subsection{Chiral decomposition of the two-dimensional
models}\label{subsection:Splitting}

We now explain the decomposition of the two-dimensional massless $(0,S)$-models
described in
Section~\ref{subsection:2dmodels} into chiral components of the form described
in Sections
\ref{subsection:ChiralZamolodchikov}--\ref{subsection:LocalChiral}.

Given a scattering function $S \in \SFlim$, consider two copies
$\Hil_{\lef/\ri}$, $y_{\lef/\ri}(\beta),
\yd_{\lef/\ri}(\beta)$, $U_{{\lef/\ri}}$, $N_{{\lef/\ri}}$ and
$\phi_{{\lef/\ri}}$, $\phi_{{\lef/\ri}}'$ of,
respectively, the Hilbert space, Zamolodchikov operators, representation of the
affine group of $\Rl$, particle number
operators and half\/line fields discussed in
Sections~\ref{subsection:ChiralZamolodchikov},
\ref{subsection:ChiralFields}. We will use the notation $y^\#_{\lef/\ri}(\psi)'
:= U_{{\lef/\ri}}(j)
y^\#_{\lef/\ri}(\bar \psi) U_{{\lef/\ri}}(j)$.

We also introduce isometries $v_{\lef/\ri} : L^2(\Rl,d\beta) \to L^2(\Rl,
dp/|p|)$ defined by
\begin{equation*}
(v_\lef \psi)(p) := \begin{cases}
\psi(\log (-p)) &\text{if }p < 0\\
0 &\text{if }p \geq  0,
\end{cases}\qquad
(v_\ri\psi)(p) := \begin{cases}
0 &\text{if }p \leq 0\\
\psi(\log p)&\text{if }p > 0.
\end{cases}
\end{equation*}
It is clear that the map $v:\psi_1 \oplus \psi_2 \in L^2(\Rl,d\beta)\oplus
L^2(\Rl,d\beta) \mapsto v_\lef\psi_1 +
v_\ri\psi_2 \in L^2(\Rl, dp/|p|)$ is unitary.
Furthermore, to a given $f \in \Ss(\Rl^2)$ we associate functions $f_{\lef/\ri}
\in \Ss(\Rl)$ through
\begin{equation}
f_\lef(\xi) := \frac{1}{2}\int_\Rl d\xi'\, f\Big(\frac{\xi+\xi'}{2},
\frac{\xi-\xi'}{2}\Big), \quad
f_\ri(\xi) := \frac{1}{2}\int_\Rl d\xi'\, f\Big(\frac{\xi+\xi'}{2},
\frac{\xi'-\xi}{2}\Big).
\end{equation}
If $f = \partial g/\partial x_k$, with $g \in \Ss(\Rl^2)$, $k=0,1$, a
calculation using~\eqref{eq:fm},
\eqref{eq:HatTransform} shows that
\begin{equation}
f^{0\pm}(e^\beta) = \frac{(-1)^{k+1}}{\sqrt{2\pi}}\hat g_\ri^\pm(\beta), \quad
f^{0\pm}(-e^\beta) = -\frac{1}{\sqrt{2\pi}}\hat g_\lef^\pm(\beta), \qquad \beta
\in \Rl,
\end{equation}
or, equivalently,
\begin{equation}\label{eq:chiralfourier}
f^{0\pm} = -\frac{1}{\sqrt{2\pi}}\big( v_\lef \hat g_\lef^\pm+(-1)^kv_\ri \hat
g_\ri^\pm).
\end{equation}

\begin{proposition}\label{prop:splitfields}
There exists a unitary operator $V :\Hil_{\lef}\otimes \Hil_{\ri}\to \Hil_0$
such that:
\begin{enumerate}
\item \label{it:zamolodchikov}For all $\psi \in \Hil_{1}$ there holds, on
$\Dc\otimes\Dc$,
\begin{align}
V^*z^\dagger_0(v_\lef\psi)V&= y_\lef^\dagger(\psi)'\otimes 1, \label{eq:VzlV}\\
V^*z^\dagger_0(v_\ri\psi)V &= S(\infty)^{N_\lef} \otimes y_\ri^\dagger(\psi)
\label{eq:VzrV}.
\end{align}
\item \label{it:poincare}$V^*U_0(x,\theta) V = U_{\lef}(x_\lef,\theta) \otimes
U_{\ri}(x_\ri,-\theta)$, where $x_{\lef}
:= x_0 + x_1$, $x_\ri := x_0 - x_1$ are the left and right light ray components
of $x = (x_0,x_1) \in \Rl^2$.
\item \label{it:reflection}$V^*U_0(j)V=S(\infty)^{N_\lef\otimes
N_\ri}(U_{\lef}(j)\otimes U_{\ri}(j)).$
\item \label{it:fields}For every $f \in \Ss(\Rl^2)$ such that  $f = \partial
g/\partial x_k$ with $g \in \Ss(\Rl^2)$,
$k=0,1$, there holds, on $\Dc\otimes\Dc$,
\begin{align}
V^* \phi_0(f) V &= -\frac{1}{\sqrt{2\pi}}\big( \phi'_{\lef}(g_\lef)\otimes 1
+(-1)^k\, S(\infty)^{N_\lef}\otimes
\phi_{\ri}(g_\ri) \big),\label{eq:fieldleft}\\
V^* \phi'_0(f) V &= -\frac{1}{\sqrt{2\pi}}\big( \phi_{\lef}(g_\lef)\otimes
S(\infty)^{N_\ri}+(-1)^k\,1 \otimes
\phi'_{\ri}(g_\ri)\big).\label{eq:fieldright}
\end{align}
\end{enumerate}
\end{proposition}

\begin{proof}
\ref{it:zamolodchikov} Recalling that $S_0(p,q) = S(\infty) = \pm 1$ for $pq <
0$ \eqref{eq:S0}, we see that, for $\psi,
\psi' \in \Hil_1$,
\begin{align*}
\zd_0(v_\ri\psi')z^\dagger_0(v_\lef\psi) &=
S(\infty)z^\dagger_0(v_\lef\psi)\zd_0(v_\ri\psi'),\\
z_0(v_\ri\psi')z^\dagger_0(v_\lef\psi) &=
S(\infty)z^\dagger_0(v_\lef\psi)z_0(v_\ri\psi').
\end{align*}
Considering then functions $\psi_1, \dots, \psi_n, \psi'_1, \dots, \psi'_{n'},
\chi_1, \dots, \chi_m,\chi'_1, \dots,
\chi'_{m'} \in \Hil_{1}$ with $n+m = n'+m' $, one has
 \begin{equation}\begin{split}\label{eq:SplitScalarProd}
 \langle z^\dagger_0(v_\lef\psi_1)\dots &z^\dagger_0(v_\lef\psi_n)
z^\dagger_0(v_\ri \chi_1)\dots z^\dagger_0(v_\ri
\chi_m)\Omega_0, z^\dagger_0(v_\lef\psi'_1)\dots z^\dagger_0(v_\lef\psi'_{n'})
z^\dagger_0(v_\ri\chi'_1)\dots
z^\dagger_0(v_\ri\chi'_{m'})\Omega_0\rangle\\
 &=  S(\infty)^{(n+n')m}\langle z_0(\overline{v_\lef \psi'_{n'}})\dots
z_0(\overline{v_\lef
\psi'_1})z^\dagger_0(v_\lef\psi_1)\dots z^\dagger_0(v_\lef\psi_n)\Omega_0, \\
 &\phantom{=  S(\infty)^{(n+n')m}\langle}\,z_0(\overline{v_\ri\chi_m})\dots
z_0(\overline{v_\ri\chi_1})
z^\dagger_0(v_\ri \chi'_1)\dots z^\dagger_0(v_\ri\chi'_{m'})\Omega_0\rangle\\
 &= \delta_{nn'}\delta_{mm'}\langle z^\dagger_0(v_\lef\psi_1)\dots
z^\dagger_0(v_\lef\psi_n) \Omega_0,
z^\dagger_0(v_\lef\psi'_1)\dots z^\dagger_0(v_\lef\psi'_n)
\Omega_0\rangle\times\\
 &\phantom{=\delta_{nn'}\delta_{mm'}}\,\langle z^\dagger_0(v_\ri\chi_1)\dots
z^\dagger_0(v_\ri\chi_m)\Omega_0,
z^\dagger_0(v_\ri\chi'_1)\dots z^\dagger_0(v_\ri\chi'_m)\Omega_0\rangle\,,
\end{split}\end{equation}
where the second equality follows from the observation that if  $n' > n$ (and
then $m > m'$), the two vectors in the
scalar product vanish, while if $n' < n$, one gets the scalar product of two
functions of $n-n' = m'-m$ variables which
have supports where all the momenta are positive, resp. negative. As
in~\eqref{eq:ScalarProd}, we have
\begin{equation*}\begin{split}
\langle z^\dagger_0(v_\lef\psi_1)&\dots z^\dagger_0(v_\lef\psi_n)\Omega_0,
z^\dagger_0(v_\lef\psi'_1)\dots
z^\dagger_0(v_\lef\psi'_n)\Omega_0\rangle\\
&= \sum_{\pi\in\frS_n}
 \int \frac{d p_1}{|p_1|}\cdots \frac{d p_n}{|p_n|}
 \prod_{j=1}^n \big(\overline{(v_\lef \psi_j)(p_j)}
(v_\lef\psi'_j)(p_{\pi(j)})\big)
 \prod_{\substack{1\leq a<b\leq n \\ \pi(a)>\pi(b)} }
 	S_0(p_{\pi(a)}, p_{\pi(b)})\\
&= \sum_{\pi\in\frS_n}
 \int d \beta_1\cdots d \beta_n
 \prod_{j=1}^n \big(\overline{\psi_j(\beta_j)} \psi'_j(\beta_{\pi(j)})\big)
 \prod_{\substack{ 1\leq a<b\leq n \\  \pi(a)>\pi(b) }}
 	S(\beta_{\pi(b)}-\beta_{\pi(a)}),
\end{split}\end{equation*}
where the last equality follows by the variable change $p_j = -e^{\beta_j}$
and~\eqref{eq:S0}.
If we now perform the further change of variables $\gamma_j =  \beta_{\pi(j)}$
and set $\sigma = \pi^{-1}$ we obtain
\begin{equation*}\begin{split}
\langle z^\dagger_0(v_\lef\psi_1)&\dots z^\dagger_0(v_\lef\psi_n)\Omega_0,
z^\dagger_0(v_\lef\psi'_1)\dots
z^\dagger_0(v_\lef\psi'_n)\Omega_0\rangle\\
&= \sum_{\sigma\in\frS_n}
 \int d \gamma_1\cdots d \gamma_n
 \prod_{j=1}^n \big(\psi'_j(\gamma_j)\overline{\psi_j(\gamma_{\sigma(j)})} \big)
 \prod_{\substack{ 1\leq a<b\leq n \\ \sigma(a)>\sigma(b)} }
 	S(\gamma_{\sigma(a)}-\gamma_{\sigma(b)})
 \\
&=
\langle
\yd_\lef(\overline{\psi'_1})\dots\yd_\lef(\overline{\psi'_n})\Omega_\lef,
\yd_\lef(\overline{\psi_1}
)\dots\yd_\lef(\overline{\psi_n})\Omega_\lef\rangle
\\
&=
\langle
\yd_\lef(\psi_1)'\dots\yd_\lef(\psi_n)'\Omega_\lef,
\yd_\lef(\psi'_1)'\dots\yd_\lef(\psi'_n)'\Omega_\lef\rangle.
\end{split}\end{equation*}
A similar (in fact, simpler) calculation shows that
\begin{multline*}
\langle z^\dagger_0(v_\ri\chi_1)\dots z^\dagger_0(v_\ri\chi_m) \Omega_0,
z^\dagger_0(v_\ri\chi'_1)\dots
z^\dagger_0(v_\ri\chi'_m) \Omega_0\rangle \\
=  \langle
\yd_\ri(\chi_1)\dots\yd_\ri(\chi_m)\Omega_\ri,
\yd_\ri(\chi'_1)\dots\yd_\ri(\chi'_m)\Omega_\ri\rangle.
\end{multline*}
Therefore, we see that the scalar product at the beginning
of~\eqref{eq:SplitScalarProd} equals
\begin{multline*}
=\delta_{nn'}\delta_{mm'}\langle
\yd_\lef(\psi_1)'\dots\yd_\lef(\psi_n)'\Omega_\lef,
\yd_\lef(\psi'_1)'\dots\yd_\lef(\psi'_n)'\Omega_\lef\rangle\times\\
\langle
\yd_\ri(\chi_1)\dots\yd_\ri(\chi_m)\Omega_\ri,
\yd_\ri(\chi'_1)\dots\yd_\ri(\chi'_m)\Omega_\ri\rangle.
\end{multline*}
As the sets
\begin{gather*}
\{ \yd_\lef(\psi_1)'\dots \yd_\lef(\psi_n)' \Omega_\lef\otimes
\yd_\ri(\chi_1)\dots \yd_\ri(\chi_m)\Omega_\ri\,:\,
\psi_1, \dots, \chi_m \in \Hil_{1}, n, m \in \Nl_0 \} \\
\{ z^\dagger_0(v_\lef\psi_1)\dots z^\dagger_0(v_\lef\psi_n)
z^\dagger_0(v_\ri\chi_1)\dots
z^\dagger_0(v_\ri\chi_m)\Omega_0\,:\, \psi_1, \dots, \chi_m \in \Hil_{1}, n, m
\in \Nl_0 \}
\end{gather*}
are total in $\Hil_{\lef}\otimes\Hil_{\ri}$ and $\Hil_0$ respectively, the
definition
\begin{equation*}
\begin{split}
V\yd_\lef(\psi_1)'\dots \yd_\lef(\psi_n)' \Omega_\lef\otimes
\yd_\ri(\chi_1)\dots \yd_\ri(\chi_m)\Omega_\ri :=
\qquad\qquad\qquad \\
 z^\dagger_0(v_\lef\psi_1)\dots z^\dagger_0(v_\lef\psi_n)
z^\dagger_0(v_\ri\chi_1)\dots z^\dagger_0(v_\ri\chi_m)\Omega_0
\end{split}
\end{equation*}
uniquely determines a unitary operator $V:\Hil_{\lef}\otimes\Hil_{\ri} \to
\Hil_0$. Equations~\eqref{eq:VzlV},
\eqref{eq:VzrV} then follow  easily.

\ref{it:poincare} We recall that $U_0$ and $U_{{\lef/\ri}}$ are second quantized
representations, so that, thanks to the
definition of $V$, it is sufficient to consider  the action of $U_0(x,\theta)$
on $v_\lef\psi, v_\ri\chi$ for
$\psi,\chi\in \Hil_{1}$. We compute:
\begin{equation}\begin{split}
(U_0(x,\theta)v_\lef\psi)(p) &= e^{i(|p|x_0-p x_1)}(v_\lef\psi)(\cosh \theta \,p
- \sinh \theta\, |p|) \\
&= e^{-ipx_\lef} (v_\lef\psi)(e^{\theta}p) = (v_\lef
U_{\lef}(x_\lef,\theta)\psi)(p),
\end{split}\end{equation}
where the second equality follows from the sign properties  of $p \mapsto \cosh
\theta \,p - \sinh \theta\, |p|$.
Similarly $U_0(x,\theta)v_\ri \chi= v_\ri U_{\ri}(x_\ri,-\theta)\chi$.

\ref{it:reflection} This also follows straightforwardly
from~\ref{it:zamolodchikov} thanks to
\begin{equation*}
U_0(j) z_0^\dagger(\varphi_1)\dots z_0^\dagger(\varphi_n)\Omega_0 =
z_0^\dagger(\overline\varphi_n)\dots
z_0^\dagger(\overline\varphi_1)\Omega_0
\end{equation*}
and similar relations for $U_{{\lef/\ri}}(j)$.

\ref{it:fields} Using~\ref{it:zamolodchikov}, equation~\eqref{eq:fieldleft}
follows by easy computations from the
definition of the fields $\phi_0$, $\phi_{\lef/\ri}$ and
equation~\eqref{eq:chiralfourier}, while
equation~\eqref{eq:fieldright} is a consequence of~\eqref{eq:fieldleft},
of~\ref{it:reflection} and of the fact that
\begin{equation*}\begin{split}
S(\infty)^{N_\lef\otimes N_\ri}(\phi_{\lef}(g_\lef)\otimes
1)S(\infty)^{N_\lef\otimes N_\ri} &=
\phi_{\lef}(g_\lef)\otimes S(\infty)^{N_\ri},\\
S(\infty)^{N_\lef\otimes N_\ri}(S(\infty)^{N_\lef}\otimes
\phi'_{\ri}(g_\ri))S(\infty)^{N_\lef\otimes N_\ri} &= 1\otimes
\phi'_{\ri}(g_\ri).
\end{split}\end{equation*}
\end{proof}

Since $\Hil_{0,1}$ is unitarily equivalent to $\Hil_{\lef,1}\oplus\Hil_{\ri,1}$,
the above result can
be seen as a generalization to the $S_0$-symmetric Fock space of the classical
result on the
tensor product decomposition of the symmetric or antisymmetric Fock space built
over a direct sum
single particle space.

Proposition~\ref{prop:splitfields}~\ref{it:fields}, together with the
half\/line-locality of the chiral fields, entails
in particular that the fields $\phi_0$, $\phi_0'$ are wedge-local. That is, we
have proved the commutation relation
\eqref{eq:PhimPhimCommutator} for the case $m=0$.
\\
\\
We now come to the decomposition of operators on $\Hil_0$. For an operator $A
\in \Ba(\Hil_{\ri})$ we define its
even/odd parts as
\begin{equation}
A_{\even/\odd} := \frac{1}{2}(A \pm S(\infty)^{N_\ri}AS(\infty)^{N_\ri}),
\end{equation}
and similarly for $A \in \Ba(\Hil_{\lef})$. Given then von Neumann algebras
$\R_{\lef/\ri}$ on $\Hil_{{\lef/\ri}}$ such
that $S(\infty)^{N_{\lef/\ri}} \R_{\lef/\ri} S(\infty)^{N_{\lef/\ri}} =
\R_{\lef/\ri}$, we consider the following
twisted tensor product von Neumann algebras:
\begin{align}
\R_\lef \hat \otimes \R_\ri &:= \R_\lef\otimes \R_{\ri,\even} +
S(\infty)^{N_\lef}\R_\lef\otimes \R_{\ri,\odd},\\
\R_\lef \check \otimes \R_\ri &:= \R_{\lef,\even}\otimes \R_\ri +
\R_{\lef,\odd}\otimes S(\infty)^{N_\ri}\R_\ri.
\end{align}
Of course, if $S(\infty)=1$, then $\R_\lef \hat \otimes \R_\ri = \R_\lef \check
\otimes \R_\ri = \R_\lef \otimes
\R_\ri$, the usual tensor product von Neumann algebras. It can be
shown~\cite{Roberts:1970} that
\begin{equation}\label{eq:twistedcommutant}
(\R_\lef \check \otimes \R_\ri)' = (\R_\lef)' \hat \otimes (\R_\ri)'.
\end{equation}

The following result will be useful in discussing the splitting of double cone
algebras of
the two-dimensional theory in the case $S(\infty)=-1$; the analogue for
$S(\infty)=1$ is trivial.

\begin{lemma}\label{lem:TwistedTensorIntersection}
Let $\R^{(i)}_{\lef/\ri} \subset \Ba(\Hil_{\lef/\ri})$, $i=1,2$, be von Neumann
algebras such that
\begin{equation*}
	(-1)^{N_{\lef/\ri}} \R^{(i)}_{\lef/\ri} (-1)^{N_{\lef/\ri}} =
\R^{(i)}_{\lef/\ri},
\end{equation*}
and define $\R_{\lef/\ri} := \R^{(1)}_{\lef/\ri}\cap\R^{(2)}_{\lef/\ri}$,
$\bar{\R}_\lef:=(-1)^{N_\lef}\R_\lef^{(1)}\cap\R_\lef^{(2)}$,
$\bar{\R}_\ri:=\R_\ri^{(1)}\cap(-1)^{N_\ri}\R_\ri^{(2)}$, $\R :=
(\R^{(1)}_\lef\hat\otimes\R^{(1)}_\ri)\cap(\R^{(2)}_\lef\check\otimes\R^{(2)}
_\ri)$. If
$\R_{\lef/\ri}$ and $\bar{\R}_{\lef/\ri}$ have trivial odd and even parts,
respectively, then $\R =
\R_\lef\otimes\R_\ri+\bar{\R}_\lef\otimes\bar{\R}_\ri$.
\end{lemma}

\begin{proof}
The von Neumann algebra $\R$ can be decomposed as $\R =
\R_{\even,\even}+\R_{\even,\odd}+\R_{\odd,\even}+\R_{\odd,\odd}$
where, denoting by $[\cdot,\cdot]_\even$ the commutator and by
$[\cdot,\cdot]_\odd$ the anticommutator,
\begin{equation}
\R_{i,j} = \{A \in \R \,:\,[(-1)^{N_\lef}\otimes 1,A]_i = 0 = [1\otimes
(-1)^{N_\ri},A]_j\,\}, \qquad i,j=\even,\odd.
\end{equation}
Similarly, defining $\R^{(1)} := \R^{(1)}_\lef\hat\otimes\R^{(1)}_\ri$,
$\R^{(2)} :=
\R^{(2)}_\lef\check\otimes\R^{(2)}_\ri$, one has $\R^{(1)} = \R^{(1)}_\even
+\R^{(1)}_\odd$, $\R^{(2)} =\R^{(2)}_\even
+\R^{(2)}_\odd$ with respect to the action of $1\otimes (-1)^{N_\ri}$ and
$(-1)^{N_\lef}\otimes 1$ respectively. It is then clear that $\R_{i,j} =
\R^{(1)}_j \cap
\R^{(2)}_i$ for $i,j =\even,\odd$. In particular,
\begin{equation*}
\R_{\even,\even}= (\R^{(1)}_\lef \otimes \R^{(1)}_{\ri,\even})\cap
(\R^{(2)}_{\lef,\even} \otimes \R^{(2)}_\ri) =
\R_{\lef,\even}\otimes \R_{\ri,\even} = \R_\lef\otimes\R_\ri.
\end{equation*}
Similarly, $\R_{\odd,\odd}=\bar{\R}_\lef\otimes\bar{\R}_\ri$. In order to get
the statement, it is
therefore sufficient to show that $\R_{\even,\odd} = \emptyset=\R_{\odd,\even}$.
To this end, consider the Tomiyama slice map $E_\lef^\omega :
\Ba(\Hil_\lef\otimes\Hil_\ri) \to \Ba(\Hil_\lef)$, $\omega
\in \Ba(\Hil_\ri)_*$, defined by the fact that $\varphi(E_\lef^\omega(A)) =
(\varphi\otimes\omega)(A)$ for all $\varphi
\in \Ba(\Hil_\lef)_*$, $A \in \Ba(\Hil_\lef\otimes\Hil_\ri)$ \cite{Sakai:1971}.
It is then easy to see that if $A \in
\R_{\odd,\even} = (\R^{(1)}_\lef \otimes \R^{(1)}_{\ri,\even})\cap
(\R^{(2)}_{\lef,\odd} \otimes
S(\infty)^{N_\ri}\R^{(2)}_\ri)$ then $E^\omega_\lef(A) \in \R_{\lef,\odd}$ and
therefore $ \R_{\odd,\even} = \emptyset$
by hypothesis. Similarly one shows that $\R_{\even,\odd} = \emptyset$.
\end{proof}

Given bounded open intervals $I, J \subset \Rl$ we introduce the double cone
\begin{equation}
O_{I,J} := \{ x \in \Rl^2 \,:\, x_\lef \in I, x_\ri \in J \}.
\end{equation}

\begin{proposition}
With $V$ the unitary of Proposition~\ref{prop:splitfields}, and with
$\M_{\lef/\ri}, \M'_{\lef/\ri}$ the von Neumann
algebras generated by the fields $\phi_{\lef/\ri}(f), \phi'_{\lef/\ri}(g)$ with
$\supp f \subset \Rl_+$, $\supp g
\subset \Rl_-$ respectively, there holds:
\begin{align}
V^*\M_0'V &= \M'_\lef \hat\otimes \M_\ri, \label{eq:splitleft} \\
V^*\M_0V &=  \M_\lef \check\otimes \M'_\ri, \label{eq:splitright}\\
V^*\A_0(O_{I,J})V
&= \A_{\lef}(I) \otimes \A_{\ri}(J) + \bar{\A}_\lef(I)\otimes\bar{\A}_\ri(J),
\label{eq:splitdoublecone}
\end{align}
\end{proposition}
where $\bar{\A}_{\lef/\ri}(a,b):=\alpha_a^{\lef/\ri}(\M_{\lef/\ri})\cap
S(\infty)^{N_{\lef/\ri}}\alpha_b^{\lef/\ri}(\M'_{\lef/\ri})$, and
$\alpha_\xi^{\lef/\ri}={\rm
Ad}\,U_{\lef/\ri}(\xi)$.

\begin{proof}
We start by showing $V^*\M_0'V \subset \M'_\lef \hat\otimes \M_\ri$. First,
observe that if $f = \partial g/\partial
x_1$ with $g \in \Ss(\Rl^2)$, thanks to the fact that the spaces of finite
particle vectors $\Dc_{\lef/\ri} \subset
\Hil_{\lef/\ri}$ are cores for $\phi'_\lef(g_\lef)$ and $\phi_\ri(g_\ri)$,
$\Dc_0$ is a core for $\phi_0(f)$ and
$V\Dc_\lef\otimes\Dc_\ri = \Dc_0$, it follows from~\eqref{eq:fieldleft} that
\begin{equation*}
V^*e^{i\sqrt{2\pi}\phi_0(f)}V = e^{i(S(\infty)^{N_\lef}\otimes \phi_{\ri}(g_\ri)
 -  \phi'_{\lef}(g_\lef)\otimes 1 )}.
\end{equation*}
If now $\supp g \subset W_L$, one has $\supp g_{\lef/\ri} \subset \Rl_\mp$ and
then $e^{-i\phi'_{\lef}(g_\lef)\otimes 1}
= e^{-i \phi'_{\lef}(g_\lef) }\otimes 1 \in \M'_\lef \hat\otimes \M_\ri$.
Moreover the identity
\begin{equation}
e^{iS(\infty)^{N_\lef}\otimes\phi_{\ri}(g_\ri)} = 1 \otimes
(e^{i\phi_{\ri}(g_\ri)})_\even + S(\infty)^{N_\lef}\otimes
(e^{i\phi_{\ri}(g_\ri)})_\odd
\end{equation}
is easily verified on finite particle vectors and entails
$e^{iS(\infty)^{N_\lef}\otimes\phi_{\ri}(g_\ri)} \in  \M'_\lef
\hat\otimes \M_\ri$. The desired inclusion is then obtained with the help of the
Trotter formula
\cite[Thm.~VIII.31]{Reed:1972}
\begin{equation}
e^{i(S(\infty)^{N_\lef}\otimes \phi_{\ri}(g_\ri)  -  \phi'_{\lef}(g_\lef)\otimes
1)} = \text{s-}\lim_{n \to \infty}\big(
e^{i(S(\infty)^{N_\lef}\otimes\phi_{\ri}(g_\ri))/n}e^{-i(\phi'_{\lef}
(g_\lef)\otimes 1)/n}\big)^n,
\end{equation}
and by analogous considerations in the case $f = \partial g/\partial x_0$.
Similarly, one gets $V^*\M_0V \subset  \M_\lef \check\otimes \M'_\ri$, but then
thanks to~\eqref{eq:twistedcommutant}
there holds
\begin{equation}
V^*\M_0V \subset  \M_\lef \check\otimes \M'_\ri =(\M'_\lef \hat\otimes \M_\ri)'
\subset V^*\M_0V,
\end{equation}
which proves~\eqref{eq:splitleft} and~\eqref{eq:splitright}.

In order to show~\eqref{eq:splitdoublecone}, we first observe that, thanks to
Poincar\'e covariance, it is sufficient to
consider $I = (-a,0)$, $J = (0, a)$, $a > 0$, so that
\begin{gather*}
\A_{\lef}(I) = \alpha^\lef_{-a}(\M_\lef) \cap\M'_\lef, \quad \A_{\ri}(J) =
\M_\ri \cap \alpha^\ri_a( \M'_\ri),\\
V^* \A_0(O_{I,J})V = (\alpha^\lef_{-a}\otimes\alpha^\ri_a)(V^*\M_0V) \cap
V^*\M'_0V = \big(\alpha^\lef_{-a}(\M_\lef)
\check\otimes\alpha^\ri_a( \M'_\ri)\big)\cap\big(\M'_\lef\hat\otimes\M_\ri\big),
\end{gather*}
where we used Proposition~\ref{prop:splitfields}.\ref{it:poincare} and
formulas~\eqref{eq:splitleft},
\eqref{eq:splitright}.
According to Proposition~\ref{proposition:EvenOps}, the algebras $\A_{\lef}(I)$,
$\A_{\ri}(J)$ have trivial odd parts and if $S(\infty)=-1$, by an analogous
statement for the anticommutator, $\bar{\A}_\lef(I)$, $\bar{\A}_\ri(J)$ have
trivial even parts. They thus satisfy the assumptions of
Lemma~\ref{lem:TwistedTensorIntersection}, which
yields~\eqref{eq:splitdoublecone}.
\end{proof}

This result completely clarifies the split of the massless two-dimensional
models into chiral theories, and the
influence of the scattering function on this decomposition. We will therefore
restrict attention to the chiral theories
on the light ray from now on.

\section{Local observables and conformal symmetry}\label{section:Conformal}

The local net $\A$ on the real line, as constructed in
Sec.~\ref{section:ChiralModels}, is covariant under the affine
group $G$, containing translations and dilations of the light ray. It is a
natural question to ask whether this model
can be extended to a conformal field theory; that is, whether the net $\A$ can
be extended to the one-point
compactification of $\Rl$ (the circle), covariant under an extension of the
representation $U$ to the M\"obius group
$\mathrm{PSL}(2,\Rl)\supset G$.

The existence of such a conformal extension is a nontrivial question. In the
physics literature, conformal symmetry is
usually derived from translation-dilation symmetry under the additional (and
sometimes implicit) assumption of existence
of a local energy density. In our context, however, the energy density is not at
our disposal. Without such additional
data, dilation symmetry does in general \emph{not} imply conformal symmetry;
counterexamples have been constructed
\cite{BuchholzSchulz-Mirbach:1990}.
Thus, we need to exploit other specific properties of the situation at hand in
order to obtain conformal extensions.

To that end, we first construct a subspace $\Hil_{\loc}\subset\Hil$ on which the
vacuum is cyclic for the local algebras
$\A(I)$.

\begin{lemma} \label{lemma:Hloc}
The subspace $\Hil_\loc := \overline{\A(a,b) \Omega} \subset \Hil$ is
independent of $-\infty \leq a < b \leq \infty$, and invariant
under $U$.
\end{lemma}
\begin{proof}
Given $0<b<\infty$, we will first show
$\overline{\A(0,b)\Omega}=\overline{\A(\Rl)\Omega}$. Let $\Psi \perp
\A(0,b)\Omega$. For any $A \in \A(0,b) \subset \M$, we know
that $A\Omega \in \mathrm{dom} \,\Delta^{1/2}$, where $\Delta^{it}$ is the
modular group
of $(\M,\Omega)$ as before. Thus the function $t \mapsto
\langle\Psi,\,\Delta^{it} A \Omega\rangle$ has an analytic continuation to the
strip $\Strip(-\frac{1}{2},0)$. But since $\Delta^{it}$ acts as a dilation
(Theorem \ref{Thm:M} c)),
the function vanishes on the boundary for $t<0$, and hence everywhere. This
shows $\Psi \perp \A(0,b')\Omega$ for any $0<b'<\infty$. Applying a similar
Reeh-Schlieder type argument to the function
$\xi \mapsto \langle\Psi\,,U(\xi) A \Omega\rangle$, using the positivity of the
generator of the translation group, we see
that $\Psi \perp \A(I)\Omega$ for any finite interval $I\subset\Rl$. Hence $\Psi
\perp \A(\Rl)\Omega$, and we arrive at
$(\A(0,b)\Om)^\perp \subset (\A(\Rl)\Om)^\perp$. But since $(\A(\Rl)\Om)^\perp
\subset (\A(0,b)\Om)^\perp$ by isotony,
$\overline{\A(0,b)\Om}=\overline{\A(\Rl)\Om}$ follows. The latter space is
invariant under $U$ by construction of
$\A(\Rl)$, which implies the lemma.
\end{proof}

The reason for considering $\Hil_\loc$ is that it is the largest space on which
we can expect an extension of $\A$ to a
net on the circle, and consequently of $U$ to the M\"obius group. Namely, if the
$\A(I)$ are covariant under such an
extension of $U$, one shows by the same methods as above that $\Hil_\loc$ is
invariant under the extended representation
as well; thus the extended net $\A$ acts on $\Hil_\loc$ with cyclic vacuum
vector.

It is a noteworthy fact that, after restriction of our net to $\Hil_\loc$, such
a conformal extension \emph{always}
exists, as we shall show now.

\begin{theorem} \label{theorem:ConformalExtension}
 The representation $U \lceil \Hil_\loc$ extends to a strongly continuous
 unitary representation of $\mathrm{PSL}(2,\Rl)$ on $\Hil_{\loc}$, and
$I\mapsto\A(I)\lceil{\Hil_\loc}$ extends to a
 local net on the circle, conformally covariant under this representation.
\end{theorem}
\begin{proof}
By construction, $\Omega$ is cyclic and separating for $\A(\Rl_+)\lceil
\Hil_{\loc}$. In fact, the modular group
associated with this pair is $\Delta^{it}\lceil \Hil_{\loc}$, since the modular
KMS condition is preserved under the
restriction. Hence the translation-dilation covariant net $\A \lceil \Hil_\loc$
has the Bisognano-Wichmann property.
Making use of the modular group of the interval algebra $(\A(0,1) \lceil
\Hil_\loc,\Om)$, the extensions of the net and
symmetry group now follow from \cite[Thm.~1.4]{GLW:extensions}.
\end{proof}

Thus, the questions of conformal symmetry and the size of the local algebras are
intimately connected: The algebras
$\A(a,b)$ are large if, and only if, the domain $\Hil_\loc$ of the extended
representation $U$ is large. In particular,
for the case $S(\infty)=-1$, Proposition~\ref{proposition:EvenOps} already gives
us a restriction: All local operators
are even. This directly implies:
\begin{proposition} \label{proposition:HLocalEven}
  If $S(\infty)=-1$, then $\Hil_\loc \subset \Hil_\even$, where $\Hil_\even$ is
the space of even particle number
vectors.
\end{proposition}

At this point, it is unknown (for general scattering function $S$) what the
actual size of $\Hil_\loc$ is; we cannot
exclude that it contains just multiples of $\Omega$, and consequently $\A(I)=\Cl
\boldsymbol{1}$. In Section
\ref{section:IsingModel}, we will however determine $\Hil_\loc$ and the local
algebras $\A(I)$ explicitly in simple
examples of $S$.

\section{Conformal scaling limits for constant scattering
functions}\label{section:IsingModel}

In this section, we illustrate the structure of the local algebras $\A(I)$, and
of their extension to a conformally
covariant theory on $\Hil_\loc$, in the examples of a constant scattering
function: $S=\pm 1$.

The simplest possible case is $S=1$. In this case, the Zamolodchikov-Faddeev
relations
\eqref{eq:zfy-1}--\eqref{eq:zfy-3} are the usual canonical commutation relations
for annihilators and creators. In fact,
one checks from the definitions that the field $\phi$ is nothing else than the
free $U(1)$ current, and $\A(I)$ the
associated local algebras; see for example \cite{BuchholzMackTodorov:1988}. It
is well known that the vacuum is cyclic
for these algebras; thus $\Hil_\mathrm{loc}=\Hil$. In fact, the representation
$U$ extends to the well-known
representation of the conformal group with central charge $c=1$.

The first non-trivial example is $S=-1$. A Euclidean version of the associated
massive two-dimensional quantum field
theory can be obtained by considering the scaling limit of the spin correlation
functions of the two-dimensional Ising
model off the critical point \cite{McCoyTracyWu:1977}. In the context of
factorizing S-matrices, this quantum field
theoretic model, and in particular its formulation on Minkowski space, is often
just referred to as ``the Ising model''.

This model has been investigated from a number of different perspectives. In
\cite{SchroerTruong:1978} and previous work
cited therein, Schroer and Truong give formulas for associated quantum field
operators. In
\cite{BergKarowskiWeisz:1979}, the form factors of one of these fields are
calculated, see also
\cite{BabujianKarowski:2004} for the calculation of the scaling dimension of
field operators in the short distance
limit. In \cite{Lechner:2005}, the existence of local observables in the
two-dimensional model, as formulated here in
terms of wedge algebras, was proven, and in \cite{BuchholzSummers:2007}, the
model was generalized to higher dimensions,
and its local and non-local aspects were discussed.

In our context, we are dealing with (a chiral component of) the massless limit
of this system, which should hence be
related to the Ising model \emph{at} the critical point. On the field
theoretical side, one expects this to be described
by a chiral Fermi field, covariant under a representation of the M\"obius group
with central charge $c=\frac{1}{2}$
\cite{MackSchomerus:1990}. In our context, it is not immediately evident that
the algebras $\A(I)$ consist of the
observables related to a Fermi field. However, we shall show now that this is
indeed the case.

On the technical side, in the case $S=-1$, our relations
\eqref{eq:zfy-1}--\eqref{eq:zfy-3} are canonical
\emph{anti}commutation relations.
As a consequence, the ``smeared'' creation and annihilation operators
$\yd(\psi)$, $y(\psi)$ are bounded, namely
\cite[Prop.~5.2.2]{BraRob:qsm2}
\begin{equation}\label{eq:CAREstimate}
   \lVert \yd(\psi) \rVert =
   \lVert y(\psi) \rVert =
   \lVert \psi \rVert_{\Hil_{1}}.
\end{equation}
This will simplify our arguments considerably.

Proposition~\ref{proposition:EvenOps} gives only an ``upper estimate'' for the
size of the local algebras $\A(I)$. We
now want to determine the size of $\A(I)$ explicitly. In fact, we will show in
detail how these algebras are generated
by the energy density of a chiral Fermi field.

To that end, it is very helpful to introduce a new field operator $\psi$ by
\begin{equation}\label{eq:PsiDef}
 \psi(\xi)
 :=
 \frac{1}{\sqrt{2\pi}}
 \int d\beta\,e^{\beta/2}
\left(\sqrt{i}\,e^{i e^\beta \xi}\yd(\beta)+\frac{1}{\sqrt{i}}e^{-i e^\beta
\xi}y(\beta)
 \right)\,.
\end{equation}
The smeared field $\psi(f)$ is a well-defined bounded operator for any test
function $f\in\Ss(\Rl)$, since the functions
$\beta \mapsto e^{-\beta/2}\hat f^\pm(\beta)$ belong to $L^2(\Rl)$. One readily
checks that $\psi(f)$ is  selfadjoint
for real $f$, and transforms covariantly under translations and dilations
according to
\begin{equation*}
 U(\xi',\la)\psi(\xi)U(\xi',\la)^{-1}
 =
 e^{\la/2}\psi(e^\la(\xi+\xi'))
 \,.
\end{equation*}
In particular, $\psi$ has scaling dimension $\frac{1}{2}$.

Using techniques similar to those in \cite[Lemma 6.1]{BuchholzSummers:2007}, we
can clarify the relation between $\psi$
and the half\/line-local fields $\phi,\phi'$.
\begin{proposition}\label{prop:psi}
 Let $a<b$, and consider test functions $f\in\Ss(a,b)$, $ g\in \Ss (b,\infty)$,
$h\in \Ss (-\infty,a)$. Then
 \begin{align}\label{eq:psi-j-comm}
 \{\psi(f),\,\phi(g)\}=0
 \,,\qquad
 \lbrack\psi(f),\,\phi'(h)\rbrack=0
 \,.
 \end{align}
The algebra $\PG_\even(a,b)$ of even polynomials in $\psi$, smeared with test
functions having support in $(a,b)$, is a
subalgebra of $\A(a,b)$, and we have
$\overline{\PG_\even(a,b)\Om}=\Hil_\even=\Hil_\loc$. The algebra
$\PG_\even(\Rl)$
acts irreducibly on $\Hil_\even$.
\end{proposition}
\begin{proof}
From the definitions \eqref{eq:ChiralFieldDef} and \eqref{eq:PsiDef}, we see
that we have $\psi(f)=\phi(k)$ if the
function $k$ fulfills
$\hat k^\pm(\beta) = i^{\mp 1/2} e^{-\beta/2}\hat f^\pm(\beta)$.
A short computation shows that such a function $k$ can in fact be found, namely
$k = K \ast f$, where $K$ is the inverse
Fourier transform of the distribution $p \mapsto 1/\sqrt{ i\, (p+i0)}$.

Due to its analyticity and boundedness properties in Fourier space, $K$ has
support in the right half line (\cite[Thm.
IX.16]{ReedSimon:1975-2}, see also \cite[Lemma 6.1]{BuchholzSummers:2007}; note
that we use different conventions for
the Fourier transform). Thus $\supp k \subset (a,\infty)$. From the relative
locality of $\phi$ and $\phi'$, see
Prop.~\ref{proposition:ChiralFields}~d), it follows that
$\lbrack\psi(f),\,\phi'(h)\rbrack =
\lbrack\phi(k),\,\phi'(h)\rbrack=0$.

To establish the first relation in \eqref{eq:psi-j-comm}, we compute in the
sense of distributions,
\begin{equation*}
\begin{aligned}
 \{\psi(\xi),\,\phi(\xi')\}
 &=
\frac{i}{2\pi}
\int d\beta \,e^{3\beta/2}
  \left(
   - \sqrt{i}\,e^{ie^\beta (\xi-\xi')}+\frac{1}{\sqrt{i}}\,e^{-ie^\beta(\xi-\xi')}
  \right)
\\
 &=
 - \frac{i^{3/2}}{2\pi}\int_{-\infty}^\infty dp\,\sqrt{p+i0}\,e^{-ip(\xi'-\xi)}
 \,.
\end{aligned}
\end{equation*}
As before, this distribution has support only for $\xi'-\xi>0$, as desired.

Now due to \eqref{eq:psi-j-comm}, even polynomials in the field $\psi$, smeared
with test functions having support in
the interval $(a,b)$, {\em commute} with both $\phi(g)$ and $\phi'(h)$. Since
all fields involved are bounded operators,
this directly implies that any such polynomial is an element of  $\A(a,b)$; see
Eq.~\eqref{eq:ChiralAlgebrasIntersect}.

For the proof of the cyclicity statement, let $\PG(a,b)$ denote the algebra of
all (even and odd) polynomials in $\psi$,
smeared with test functions supported in $(a,b)$. Then $\Omega$ is cyclic for
$\PG(a,b)$. (This follows with arguments
as in Prop.~\ref{proposition:ChiralFields}~\ref{it:ReehSchlieder}; the extra
factor $e^{\beta/2}$ in \eqref{eq:PsiDef}
can be absorbed in the test functions.) Applying the projector
$E_\even=\frac{1}{2}(1+(-1)^N)$ onto $\Hil_\even$, we
obtain
\begin{equation*}
 \Hil_\even
 =
 E_\even\Hil
 =
 E_\even \overline{\PG(a,b)\Om}
  =
 \overline{E_\even\PG(a,b)E_\even\Om}
 =
 \overline{\PG_\even(a,b)\Om}
 \,.
\end{equation*}
Further,
\begin{equation*}
 \overline{\PG_\even(a,b)\Om} \subset \overline{\A(a,b)\Om} = \Hil_\loc.
\end{equation*}
But from Prop.~\ref{proposition:HLocalEven}, we know that $\Hil_\loc \subset
\Hil_\even$. Hence $\Hil_\even = \Hil_\loc
= \overline{\PG_\even(a,b)\Om}$.

Irreducibility of $\PG_\even(\Rl)$ now follows from cyclicity of $\Omega$ and
from the spectrum condition for
translations by standard arguments \cite[Theorem~4-5]{StrWig:PCT}.
\end{proof}

Having seen that the even local algebras $\A(I)$ are non-trivial, we now want to
understand their structure more
explicitly in terms of local field operators. To this end, we first note that
$\psi$ satisfies the anticommutation
relation of a free Fermi field,
 \begin{align}\label{eq:psi-comm}
    \{\psi(\xi),\,\psi(\xi')\}
 =
\frac{1}{2\pi}\int d\beta\,e^\beta
\left( e^{i e^\beta (\xi-\xi')}+e^{-i e^\beta (\xi-\xi')}\right)
 =
 \delta(\xi-\xi')
 \,.
 \end{align}
 This observation suggests to introduce a normal ordered even field,
\begin{equation}\label{eq:TDef}
 T(\xi)
 :=
 \frac{i}{2}\,:\!\psi(\xi)\partial_x\psi(\xi):\;
 =
 \frac{i}{2}\,\lim_{\xi'\to \xi}\left(
\psi(\xi)\partial_{\xi'}\psi(\xi')-\hrskp{\Om}{\psi(\xi)\partial_{\xi'}\psi(\xi')\Om}\right)
,
\end{equation}
as a candidate for a local energy density. This limit exists in the sense of
matrix elements between vectors from
$\DD_0$, where $\DD_0 \subset \DD$ denotes those vectors in which each
$n$-particle component is smooth and of compact
support. Expressing $T(\xi)$ in terms of creation and annihilation operators, see
Eq.~\eqref{eq:T-concrete} below, it is
also easy to see that $T$ is an operator-valued distribution.
\pagebreak
\begin{proposition}\label{prop:T}
The field $T$ is point-local, relatively local to the algebras $\PG_\even(a,b)$,
transforms covariantly under $U$, and
integrates to the generator $H$ of translations,
\begin{align}\label{T-P}
 \int_{-\infty}^\infty d\xi\,T(\xi)
 =
 H
 \,,
\end{align}
where the integral is understood in the sense of matrix elements between vectors
from $\DD_0$.
With central charge $c=\frac{1}{2}$, we have the L\"uscher-Mack commutation
relations,
\begin{align}\label{eq:T-comm}
 i[T(\xi),T(\xi')]
 &=
 -\delta'(\xi-\xi')\,(T(\xi)+T(\xi'))+\frac{c}{24\pi}\,\delta'''(\xi-\xi')
 \,.
\end{align}
\end{proposition}
\begin{proof}
In view of the anticommutation relation \eqref{eq:psi-comm}, we also have
$\{\psi(\xi),\,\partial_{\xi'}\psi(\xi')\}=-\delta'(\xi-\xi')\cdot1$ and
$\{\partial_\xi\psi(\xi),\,\partial_{\xi'}\psi(\xi')\}=-\delta''(\xi-\xi')\cdot1$, which
implies that $T$ is a point-local field.
Relative locality to $\PG_\even$ follows from \eqref{eq:psi-comm} as well. The
covariance of $T$ under translations and
dilations is clear from its definition; note that $T$ has scaling dimension two.

To establish \eqref{T-P}, we write down the normal ordered product
\eqref{eq:TDef} in terms of creation and annihilation
operators,
\begin{equation}
    \label{eq:T-concrete}
\begin{aligned}
 T(\xi)
 &=
 -\frac{1}{4\pi}\int d\beta \int d\gamma \, e^{(\beta+3\gamma)/2}
  \Big(
    i\,e^{i( e^\beta +e^{\gamma} )\xi}\,\yd(\beta)\yd(\gamma)
    +i\,e^{-i(e^\beta +e^{\gamma})\xi}\,y(\beta)y(\gamma)
    \\
    &\qquad\qquad\qquad\qquad
    -e^{i(e^\beta -e^{\gamma})\xi}\,\yd(\beta)y(\gamma)
    -e^{-i(e^\beta -e^{\gamma})\xi}\,\yd(\gamma)y(\beta)
    \Big)
    \,.
\end{aligned}
\end{equation}
(We read this in the sense of sesquilinear forms on $\DD_0 \times \DD_0$.) The
first two terms, containing two creators
and annihilators, respectively, vanish after integration over $\xi$, because they
involve exponentials of $\pm
i(e^\beta+e^{\gamma})\xi$ and the factor $(e^\beta+e^{\gamma})$ is strictly
positive. Therefore,
\begin{equation*}
\begin{aligned}
 \int_{-\infty}^\infty d\xi\,T(\xi)
 &=
 \frac{1}{4\pi}
 \int_{-\infty}^\infty d\xi
 \int d\beta \,d\gamma\,e^{(\beta+3\gamma)/2}
  \Big(
    e^{i(e^\beta -e^{\gamma})\xi}
    +e^{-i(e^\beta -e^{\gamma})\xi}
    \Big)\yd(\gamma)y(\beta)
  \\
  &=
  \int d\beta \, e^\beta
\,
\yd(\beta)y(\beta)
=
H
\,.
\end{aligned}
\end{equation*}

In summary, $T$ is a local, translation and dilation covariant field of scaling
dimension two that integrates up to $H$
and is relatively local to the net $\PG_\even$, which acts irreducibly on
$\Hil_\even$ (Prop.~\ref{prop:psi}). Hence the
hypotheses of the L\"uscher-Mack theorem \cite{LuescherMack:1976} are fulfilled
(see
\cite[Thm.~3.1]{FurlanSotkovTodorov:1989}), and the commutation relation
\eqref{eq:T-comm} follows. The value of the
central charge $c$ can then be computed from the vacuum two-point function.
\begin{equation*}
\begin{aligned}
 \hrskp{\Om}{T(\xi)T(\xi')\Om}
 &=
 -\frac{1}{16\pi^2}\int d\beta_1 \int d\beta_2 \int d\gamma_1 \int d\gamma_2
 \,e^{(\beta_1+\beta_2+3\gamma_1+3\gamma_2)/2}
 \\
 &\qquad
 \times \exp\big(-i(e^{\beta_1} + e^{\gamma_1} )\xi
  +i(e^{\beta_2} + e^{\gamma_2} ) \xi' \big)
  \\ &\qquad
 \times \hrskp{\Om}{y(\beta_1)y(\gamma_1)\yd(\beta_2)\yd(\gamma_2)\Om}
 \\
 &=
 \frac{1}{2}\cdot\frac{1}{24\pi}\cdot\frac{1}{2\pi}\int_0^\infty
dk\,k^3\,e^{-ik(\xi-\xi')}
 \,.
\end{aligned}
\end{equation*}
Taking the antisymmetric part of this distribution and comparing it with
\eqref{eq:T-comm}, we read off $c=\frac{1}{2}$.
\end{proof}

This identifies the field $T$ as  the energy density of a real free chiral Fermi
field. Expanding $T$ into its Fourier
modes $L_n$, $n\in\Zl$, we therefore have a representation of the Virasoro
algebra with central charge $\frac{1}{2}$ in
our chiral net $\A$, and a corresponding subnet $I\mapsto{\rm Vir}_{1/2}(I)$ on
$\Hil_\even$. This net transforms
covariantly under a unitary representation of the M\"obius group, with the
generator $K$ of the special conformal
transformations given by $K=\int d\xi\,\xi^2 \, T(\xi)$
\cite{FurlanSotkovTodorov:1989}.

The local algebras are now completely fixed by the following theorem.

\begin{theorem} \label{theorem:IsingAlgebras}
 In the chiral model with scattering function $S=-1$, the net of local von
Neumann algebras is the Virasoro net with
central charge $\frac{1}{2}$, i.e., $\A(I)={\rm Vir}_{1/2}(I)$ for any interval
$I\subset\Rl$.
\end{theorem}
\begin{proof}
Both nets, $\A$ and ${\rm Vir}_{1/2}$, can be restricted to the even subspace
$\Hil_\loc=\Hil_\even\subset\Hil$, and
both have the vacuum as a cyclic vector on this space -- see
Proposition~\ref{prop:psi} regarding cyclicity for
$\PG_\even(I) \subset {\rm Vir}_{1/2}(I)$. For every interval $I$, we have ${\rm
Vir}_{1/2}(I)\subset\A(I)$ by
construction, and the same then follows for any subset $I \subset \Rl$,
cf.~\eqref{eq:AlgGenSubset}. But the Virasoro
net on the real line is Haag-dual \cite{KawahigashiLongo:2004}. Hence
\begin{equation*}
{\rm Vir}_{1/2}(I)
\subset \A(I)
\subset \A(I')'\subset
{\rm Vir}_{1/2}(I')'
={\rm Vir}_{1/2}(I)\,,
\end{equation*}
which implies ${\rm Vir}_{1/2}(I)=\A(I)$.
\end{proof}

\section{Conclusions}\label{section:conclusion}

In this paper, we have investigated the short-distance scaling limit of
1+1-dimensional models of quantum field theory
with a factorizing scattering matrix, for a certain class of two-particle
scattering functions $S$. At finite scale,
these models are generated by wedge-local field operators depending on $S$ in an
explicit manner. Proceeding to scale
zero, we showed that this feature is also maintained in the limit, and
investigated the limit theories in terms of their
generators.

As might heuristically be expected, the limit turned out to be a massless,
dilation covariant theory which extends (trivially, if $S(\infty)=1$) a chiral
theory. We were able to establish this fact on the
level of local von Neumann algebras: The
observable algebras $\A(O)$ associated with double cones contain the tensor
products of local interval algebras $\A(I)$ of the
chiral components. For algebras associated with unbounded regions (wedges and
half-lines), one obtains a tensor product
as well, but with a grading in the case $S(\infty)=-1$.

We then investigated in more detail the individual chiral components of the
limit theory, which are of interest in their
own right. They are translation-dilation-reflection covariant models on the real
line; and while they are massless, they
are formally very similar to the massive two-dimensional models, viewed in
rapidity space. These theories can be defined
on the level of half-line algebras or, by considering intersections, on the
level of interval algebras.

Our particular approach to the scaling limit via the wedge-local fields has the
merit that the computation of the limit
is rather easy on the level of the generators, but this comes at the price of an
indirect characterization of the local
fields and observables of the limit theory. In particular, the nontriviality of
the interval algebras $\A(I)$ is not
guaranteed by our construction. The analysis presented here is thus
complementary to other approaches to the
short-distance behavior of the models considered, and it is interesting to
compare the different procedures.

In case the point-local quantum fields contained in the models at finite scale
are sufficiently explicit, one might base
the scaling limit analysis on these quantities. However, as the S-matrix is
taken here as the main input into the
construction, for most of the models no Lagrangian formulation or local fields
are known. Moreover, even if point-local
fields can be constructed, for example by Euclidean perturbation theory, their
relation to the real-time S-matrix is
very indirect. One can therefore expect a rigorous analysis of the connection
between the collision operator on the one
hand and the short-distance limit on the other hand to be quite involved with
this method. For example, in the Ising
model explicit formulas for local fields are available, but have a rather
complicated form \cite{McCoyTracyWu:1977}. By
comparison, the S-matrix and wedge-local generators of this model are extremely
simple. As we have shown, it is possible
to circumvent the construction of the local fields at finite scale, and
still completely analyze the corresponding scaling limit theory.

Because observables localized in bounded space-time regions are only
characterized indirectly in our approach, a
detailed comparison with techniques based on local observables is difficult. One
can expect however that the limit of
double-cone-local objects would possibly yield \emph{less} (but in no case more)
limit points than those obtained when
working with wedge-local objects, in some analogy to the scaling limit of charge
sectors
\cite{DMV:scaling_charges,DAnMor:supersel_models}. In this sense, the limit
theory that we compute is maximally large.

Another approach to the scaling limit is that of Buchholz and Verch
\cite{Buchholz:1995a}. Here one defines the limit in
terms of bounded local operators, and in this sense of more general objects,
since unitaries $\exp i \phi(f)$ and their
weak limit points would be included. This might yield a larger limit theory than
ours, and indeed, one expects
\cite{Buc:quarks} a large center to occur in the limit algebras. Due to
technical difficulties in fully describing this
central part of the algebras, we did not yet proceed in this direction. These
problems are present even in the free
field case, and their complete clarification will probably require a
modification of the Buchholz-Verch framework. We
hope to return to this point elsewhere. It is not excluded that such a more
general approach would yield additional
``quantum'' observables as well, not only ``classical'' observables in the
center of the algebras.  Nevertheless, let us
remark that the wedge algebras $\M_0,\widehat{\M}_0$ that we constructed in the
limit theory are Haag-dual, and to this degree maximal; any additional local
observables could only be accommodated on
an extended Hilbert space.

In the approach chosen here, a central question turned out to be whether the
chiral models constituting the scaling
limit extend to conformal quantum field theories on the circle, covariant under
the M\"obius group. We showed that there
is indeed always such an extension, namely on the subspace $\Hil_\loc \subset
\Hil$ generated from the vacuum by the
local algebras. In this sense, a conformal extension exists if and only if the
local algebras are large. As a general
feature, we showed that in the case $S(\infty)=-1$, the local subspace
$\Hil_\loc$ contains only even particle number
vectors, and all local operators must be even with respect to the particle
number as well.

This effect is illustrated by the models with the constant scattering functions
$S=\pm1$. In these two cases, we
explicitly computed the local algebras of the chiral components. For the free
field ($S=1$), one obtains the minimal
model with conformal charge $c=1$, and for the Ising model ($S=-1$), one obtains
the minimal model with $c=\frac{1}{2}$.
However, in the case of non-constant $S$, the exact size of the local algebras
remains an open question. In fact, our
present results do not rule out the possibility that they are trivial in the
sense $\Hil_\loc=\Cl\Om$.

Another related problem is to clarify the significance of the function $S$
entering in the definition of our chiral
models. At finite scale, $S$ directly corresponds to the S-matrix, and its
physical interpretation is clear
\cite{Lechner:2008}. From a more mathematical point of view, $S$ is an invariant
(under unitary equivalence) of the
two-dimensional massive nets, and in particular, two models with different
scattering function are never equivalent. By
comparison, the significance of $S$ is much less understood in the scaling
limit, despite it formally being equal to a
two-dimensional scattering function. For $S(\infty)=1$, it seems clear that any
formulation of scattering theory of massless
two-dimensional models (cf.~\cite{Buchholz:1975}) yields a trivial scattering
matrix, due to the tensor product
structure of the local algebras. In the terminology of
\cite{FendleySaleur:1994}, we deal with models with trivial
left-right scattering, while our scattering function $S$ determines the
left-left and right-right scattering.
However, on a single light ray or in a single chiral component, scattering
theory in the usual sense cannot be
formulated, and is not physically meaningful.

Furthermore, $S$ is not known to be an invariant of the chiral models. Therefore
it is possible that models with
different scattering functions, inequivalent at finite scale, become equivalent
in the scaling limit. Such an effect
would actually be expected for asymptotically free theories, and is supported by
results obtained in another approach to
massless factorizing scattering: In \cite{ZamolodchikovZamolodchikov:1992}, a
model similar to ours -- yet with a richer
particle spectrum -- is analyzed by means of a Thermodynamic Bethe Ansatz.
Assuming for a moment that the results of
\cite{ZamolodchikovZamolodchikov:1992} can be transferred to our situation by
analogy, one is lead to the conjecture
that our local net $I \mapsto \A(I)$ is actually unitarily equivalent to the
minimal conformal model with $c=1$ (for
$S(\infty)=1)$ or $c=\frac{1}{2}$ (for $S(\infty)=-1$), irrespective of the
details of the function $S$. This would mean
that the interaction described by $S$ vanishes in the scaling limit, and that
the limit models are actually complicated reparametrizations of the free Bose or
Fermi field.  However, the technical
arguments of \cite{ZamolodchikovZamolodchikov:1992} are largely based on
thermodynamical considerations and do not
directly apply in our context.

This situation would be compatible with our present results as well. A rigorous
answer to the question which of the
possible scenarios, ranging from trivial local algebras to asymptotically free
theories, is realized for which
scattering function, would deepen our understanding of the short distance
structure of quantum field theory. Further
results in this direction will be presented elsewhere.

\subsubsection*{Acknowledgements}
Work on this project originated in many joint discussions with Claudio D'Antoni.
His sudden and untimely death in
October 2010 robbed us of a friend and collaborator. We dedicate this article to
his memory.
\pagebreak

\end{document}